\documentclass[iop]{emulateapj}
\usepackage{graphicx}
\usepackage{amsmath}
\usepackage{natbib}

\begin{document}

\slugcomment{to be submitted to ApJ}
\shortauthors{Harris et al.}

\title{Globular Cluster Systems in Brightest Cluster Galaxies. II:  NGC 6166}
        
\author{ William E. Harris\altaffilmark{1}, 
         John P.~Blakeslee\altaffilmark{2},
         Bradley C.~Whitmore\altaffilmark{3}, 
         Oleg Y. Gnedin\altaffilmark{4}, 
	 Douglas Geisler\altaffilmark{5}, and
	 Barry Rothberg\altaffilmark{6}
}

\altaffiltext{1}{Department of Physics \& Astronomy, McMaster University, Hamilton, ON, Canada; harris@physics.mcmaster.ca}
\altaffiltext{2}{Herzberg Institute of Astrophysics, National Research Council of Canada, Victoria, BC V9E 2E7, Canada; jblakeslee@nrc-cnrc.gc.ca}
\altaffiltext{3}{Space Telescope Science Institute, 3700 San Martin Drive, Baltimore MD 21218, USA; whitmore@stsci.edu}
\altaffiltext{4}{Department of Astronomy, University of Michigan, Ann Arbor, MI 48109; ognedin@umich.edu}
\altaffiltext{5}{Departamento de Astronomi\'a, Universidad de Concepci\'on, Casilla 160-C, Concepci\'on, Chile; dgeisler@astroudec.cl}
\altaffiltext{6}{LBT Observatory, University of Arizona, 933 N.Cherry Ave, Tucson AZ 85721, USA; dr.barry.rothberg@gmail.com}
%\altaffiltext{6}{Department of Astronomy, Peking University, Beijing 100871, China; peng@bac.pku.edu.cn}
%\altaffiltext{7}{Department of Physics and Astronomy, University of Alabama, Box 870324, Tuscaloosa, AL 35487-0324, USA; jbailin@ua.edu}
%\altaffiltext{9}{Department of Physics and Astronomy, Western Washington University, Bellingham WA 98225, USA; Regina.BarberDeGraaff@wwu.edu}

\date{\today}

\begin{abstract}
	We present new deep photometry of the globular cluster system (GCS) around NGC 6166, 
the central supergiant galaxy in Abell 2199.  HST data from the ACS and WFC3 cameras in $F475W, F814W$ 
are used to determine the spatial distribution of the GCS, its metallicity distribution function (MDF),
and the dependence of the MDF on galactocentric radius and on GC luminosity.  The MDF is extremely 
broad, with the classic red and blue subpopulations heavily overlapped, but 
a double-Gaussian model can still formally match the MDF closely.  
The spatial distribution follows a S\'ersic-like profile detectably to a projected radius of 
at least $R_{gc} = 250$ kpc. To that radius,
the total number of clusters in the system is $N_{GC} = 39000 \pm 2000$, the global specific frequency
is $S_N = 11.2 \pm 0.6$, and 57\% of the total are blue, metal-poor clusters.
The GCS may fade smoothly into
the Intra-Cluster Medium of A2199; we see no clear transition from the core of the galaxy to the cD halo or the ICM.
The radial distribution, projected ellipticity, and mean metallicity of the red (metal-richer) clusters 
match the halo light 
extremely well for $R_{gc} \gtrsim 15$ kpc, both of them varying as $\sigma_{MRGC} \sim \sigma_{light} \sim R^{-1.8}$.  
By comparison, the blue (metal-poor) GC component has a much shallower falloff $\sigma_{MPGC} \sim R^{-1.0}$ and a more nearly
spherical distribution.
This strong difference in their density distributions produces a net metallicity gradient
in the GCS as a whole that is primarily generated by the population gradient.  
With NGC 6166 we appear to be penetrating into a regime of high enough galaxy mass and rich enough environment
that the bimodal two-phase description of GC formation is no longer as clear or effective as it has been in smaller galaxies.

\end{abstract}


\section{Introduction}

Brightest Cluster Galaxies (BCGs) are the largest galaxies in the universe, 
and as such they are likely to have evolved from the most complex and extended hierarchical-merger trees
during the most rapid stage of galaxy assembly.  
Their growth is still ongoing today as they 
accrete smaller galaxies within their host clusters.

BCGs also host the richest populations of globular clusters (GCs), a mark of 
exceptionally intense star formation under conditions of high gas density at high redshift.  The nearest
examples of these high-specific-frequency globular cluster systems (GCSs) include those within 
M87 in Virgo \citep{harris09b}, NGC 1399 in Fornax \citep{bassino2006}, and NGC 3311 in Hydra
\citep{wehner_etal08}.  These cases are, however, eclipsed by the still more luminous
giants that can be found by searching further outward.  A well known example is NGC 4874 in Coma
\citep{peng_etal11}, which may hold $\simeq 23000$ GCs of its own, and a still richer system 
may lie within Abell 1689 \citep{alamo-martinez_etal2013}.  Furthermore, a rich galaxy cluster may
also contain an extended Intra-Cluster Medium (ICM) of stellar light and high-temperature X-ray gas,
and the ICM itself can hold large numbers of intragalactic globular clusters (IGCs) that may even exceed the total in the central
BCG \citep[see][]{peng_etal11,durrell_etal2014}.  The IGCs may in turn be a combination of objects stripped from
other galaxies in the cluster, and ones in the cD halo of the central BCG.  In short, 
these systems offer a testing ground of unequalled richness
for exploring GC systematics observationally.

Incorporating GCs fully into hierarchical galaxy formation models is difficult
because spatial resolutions less than $\sim 1$ pc are needed to trace star cluster formation,
while the galaxy as a whole needs a scale six orders of magnitude larger. But
appropriately designed models have had some initial success 
at reproducing the observed GC mass distribution, and perhaps more challengingly, the metallicity distribution 
\citep[e.g.][]{kravtsov_gnedin05,muratov_gnedin10, griffen2010,tonini2013,li_gnedin14}. 
The existing models, though still quite preliminary, already hint that the GC metallicity distribution 
function (MDF) changes significantly with
host galaxy mass even among large galaxies. The BCGs represent the relatively unexplored
extreme upper limit of any such trends.

In Paper I \citep{harris_etal2014}, we introduced a new HST-based imaging survey of seven BCGs,
aimed primarily at studying the GCSs in these biggest of all galaxies.  Paper I contained
a discussion of the luminosity and mass distribution function of
their GC populations.  In the current paper, we present more detailed results
for the nearest of these seven systems, NGC 6166, including
the GCS spatial distribution and total population, and the distribution of GCs
by color and metallicity.  Similar material for the remaining six galaxies will be presented in the next paper of
our series.

For NGC 6166 we assume $d = 130$ Mpc for $H_0 = 70$ km s$^{-1}$ Mpc$^{-1}$ 
along with a foreground reddening $E(F475W-F814W) = 0.023$ at ($\ell = 62.9^o, b = 43.7^o$).
The adopted distance modulus is $(m-M)_I = 35.60$ and the galaxy luminosity is $M_V^T = -23.7$ (Paper I).

\section{Photometric Reductions}

NGC 6166 is a classic cD galaxy with an extremely extended halo \citep[][hereafter B15]{bender_etal2015}
and is the central supergiant galaxy in Abell 2199.
Early detections of its rich GCS with ground-based imaging were done by 
\citet{pritchet_harris1990} from the Canada-France-Hawaii Telescope, 
\citet{bridges_etal1996} from the William Herschel Telescope, and \citet{blakeslee_etal1997} with the 
MDM Observatory.  With ground-based
imaging, however, only the brightest magnitude or two of the GC luminosity
function and approximate mean color indices could be
measured, yielding very uncertain estimates of the spatial extent or specific frequency of the system.

\begin{figure}[t]
 \vspace{0.0cm}
 \begin{center}
\includegraphics[width=0.5\textwidth]{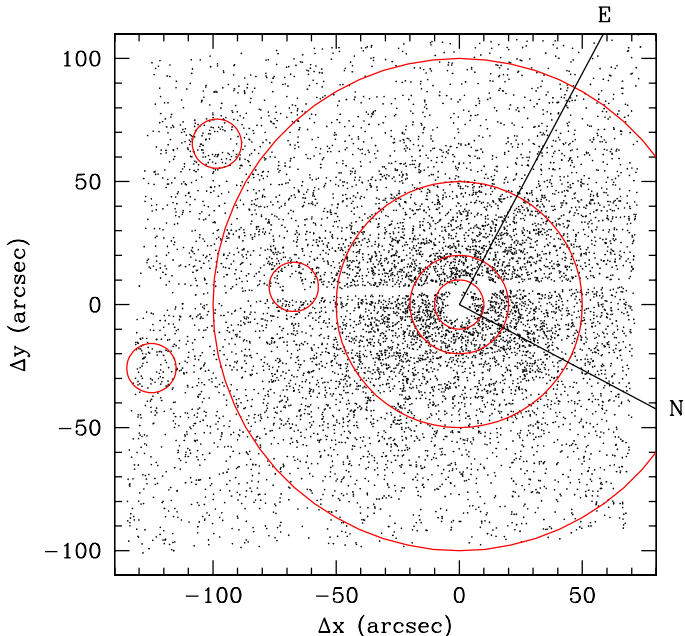}
\end{center}
\vspace{-0.5cm}
\caption{Locations of the measured starlike objects ($x,y$ in arcseconds relative
	to the center of NGC 6166) for the brightest objects, $F814W < 27.2$.
	The large concentric circles have radii of $20'', 50'', 100''$.  The three
	smaller circles of radii $10''$ mark the three companion galaxies NGC 6166A,
	NGC 6166D, and PGC058261 (see text).  The visible gap running across the
middle of the frame is the gap between the two CCD detectors in the ACS/WFC. 
{\bf The fiducial directions towards North and East on the sky are marked; the +x-axis of
the ACS camera is oriented $28^o$ E of N.} }
\vspace{0.5cm}
\label{fig:xy_acs}
\end{figure}

By contrast, the HST cameras are extremely well suited to imaging of GCSs in giant galaxies
at distances of $\sim 100-200$ Mpc where the field sizes of either the ACS or WFC3 
arrays correspond to linear diameters near 100 kpc,
while the bright half of the GC luminosity function (GCLF) can be 
well measured in just a few orbits of exposure time.  

In this paper, we present the first comprehensive 
two-color photometric study of the NGC 6166 system.
The basic design of the program is set out in Paper I.
For NGC 6166, the ACS/WFC camera was nearly
centered on the target galaxy, while the WFC3 camera was used in parallel to obtain an additional
field in the outskirts of the host cluster A2199.
Total exposure times for ACS/WFC were 5370 sec (F475W) and 4885 sec (F814W), 
while for WFC3 they were 5460 sec (F475W) and 4555 sec (F814W).
As described in Paper I, the total exposures were designed to reach at least as faint as
the expected GC luminosity function peak frequency (turnover point) 
at $M_{V,0} \simeq -7.3, M_{I,0} \simeq -8.4$, so that the bright half of the distribution
would be securely measured.

Individual GCs in all types of galaxies have typical effective diameters of $\sim 5$ pc \citep[e.g.][]{jordan_etal05,harris09a}.
Thus for galaxies at distances $d \lesssim 50$ Mpc, HST imaging will resolve many or most GCs,
and extra efforts must be made to obtain integrated magnitudes appropriately
corrected for their individual profiles and scale radii \citep[e.g.][]{mieske_etal2006,peng_etal2009,harris09a}.
However, at $d > 100$ Mpc 
almost all of the individual GCs appear
starlike:  their angular diameters will be typically $<  0.01\arcsec$, well below the
$0.1\arcsec$ resolution of the HST and thus in the
``unresolved'' category as discussed in \citet{harris09a}.  
The advantages for photometric measurement are that
the GCs can be measured through standard point-spread-function (PSF) fitting, and that
they can be easily distinguished from the great majority of the faint, nonstellar
background galaxies that constitute the main source of sample contamination.

\begin{figure}[t]
 \vspace{0.0cm}
 \begin{center}
\includegraphics[width=0.5\textwidth]{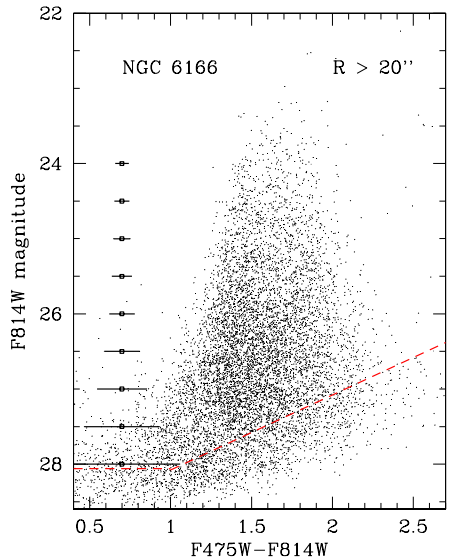}
\end{center}
\vspace{-0.5cm}
\caption{Color-magnitude diagram for the globular cluster population around
NGC 6166, showing the radial zone $20''-160''$ (equivalent to $R = 12 - 100$ kpc).
The red dashed line across the lower part of the diagram shows the 50\% detection completeness
level of the photometry, while the errorbars along the left show the measurement 
uncertainties in the color index.}
\vspace{0.5cm}
\label{fig:cmd}
\end{figure}

We started the data analysis from the $*.flc$ files provided by the HST Archive.
With \emph{stsdas/multidrizzle} a single
combined image was then generated in each filter which was CTE-corrected, mostly free of cosmic
rays, and corrected for geometric distortion.
Photometry was carried out with the standard tools in SourceExtractor
\citep[SE;][]{bertin_arnouts96} and DAOPHOT \citep{stetson87} in its IRAF
implementation, including aperture photometry (\emph{phot}) followed by
PSF fitting through \emph{allstar}.
First, the images in both filters were registered and combined to produce a master
white-light image, then SE was run on that master image to produce a very deep finding list of objects.
This list was used as input to \emph{daophot/phot} for the images in both filters.
Clearly nonstellar or crowded objects were deleted through the use of SE and \emph{allstar}
parameters:  specifically, objects were kept if within $0.9 < r_{1/2} < 1.4$ px,
$\chi < 2$, and $err < 0.3$ mag.
See \citet{harris09a} for detailed examples of the procedure.

The PSFs were empirically generated from bright, uncrowded starlike objects distributed across the target fields.
For the ACS fields, 95 stars in $F475W$ and 87 stars in $F814W$ were summed to generate
the PSFs, while for WFC3 42 stars in $F475W$ and 39 stars in $F814W$ were used.
The PSF shape was set to be quadratically variable in position $(x,y)$, though comparisons
with the uniform-PSF option showed negligible differences in the resulting photometry.
The PSF-fitted magnitudes were corrected to large-aperture magnitudes ($r = 0.5\arcsec$)
with aperture photometry of bright isolated stars, and lastly corrected to total magnitudes
with the enclosed-energy curves published in the ACS/WFC and WFC3 Handbooks.

The final data list consists of starlike objects that were measurable on both filters,
and that had $(x,y)$ positions matching between filters to within 0.1 arcsecond.
We report our results in the natural filter-based magnitudes
$F475W, F814W$ and in the VEGAMAG system. Values for the
filter zeropoints given on the HST webpages appropriate for the dates of the 
exposures have been used as follows:
$F475W_0$ = 25.778 (WFC3) or 26.154 (ACS), $F814W_0$ = 24.680 (WFC3) or 25.523 (ACS).
The color index $(F475W-F814W)$ is close to $(g-I)$, but can also be
transformed to $(B-I)$ through \citep{saha_etal11} 
\begin{eqnarray}
(B-I) \, & = & \, 1.185\, (F475W-F814W), \nonumber\\
   I  \, & = & \, F814W + 0.014\, (B-I) \, .
  \label{eq:trans}
\end{eqnarray}

In Figure \ref{fig:xy_acs}, the $xy$ locations of the 8223
objects brighter than $F814W = 27.2$ (clearly brighter than the photometric completeness limit, as discussed in the next section),
are plotted.  The field also contains some companion galaxies in A2199,
the three brightest of which are NGC 6166D (at upper left in Fig.~\ref{fig:xy_acs}), NGC 6166A
(on the lower left edge), and PGC058261 (left of center).  These are marked in the Figure with
small circles of radii $10''$.  These appear to have small GC populations of their own,
but clearly make up very minor additions to the overwhelmingly larger population around
NGC 6166 itself.  In the very center ($R < 10''$) we find the well known ``multiple nucleus''
of NGC 6166 where three small cluster galaxies lie in projection against its central bulge.
Analysis of their light profiles (see B15)
indicates that these companions are relatively undistorted and thus not physically connected
with the central BCG.

\begin{figure}[t]
 \vspace{0.0cm}
 \begin{center}
 \includegraphics[width=0.5\textwidth]{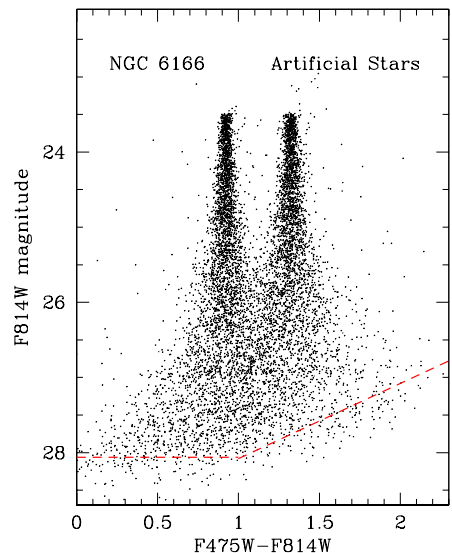}
\end{center}
\vspace{-0.5cm}
\caption{Color-magnitude diagram for artificial stars in the ACS/WFC field around
NGC 6166.  The measured stars here are ones falling in the radial zone $R > 20''$.
Fake stars were inserted with magnitudes and colors falling on two dispersionless vertical
sequences, so the measured scatter in color seen in this Figure
represents the internal measurement uncertainty.
The red dashed line across the lower part of the diagram shows the 50\% detection completeness
level of the photometry as in Fig.~\ref{fig:cmd}.}
\vspace{0.5cm}
\label{fig:fakecmd}
\end{figure}

The final color-magnitude diagram for the starlike objects in the ACS field is shown
in Figure \ref{fig:cmd}.  
It includes 11371 objects in the radial range $R > 20''$ and excludes  the $10''$ regions
around the three companion galaxies as defined in Fig.~\ref{fig:xy_acs}.
An enormous GC system is present, but the spread in color is large, and
the normal blue, metal-poor (MP) and red, metal-rich (MR) subpopulations are considerably
less distinguishable than in most other galaxies.  
We analyse this issue more carefully in the next sections.

\section{Completeness and Measurement Uncertainties}

To quantify the completeness and photometric uncertainties we carried out extensive artificial-star
tests through \emph{daophot/addstar}, independently of similar experiments done
for the luminosity-function analysis in Paper I.  In a series of trials we added
mock stars into the original images that were
designed to mimic roughly the colors of the classic `blue' and `red' globular
cluster sequences.  The input artificial stars followed
dispersionless vertical sequences separated by 
$\Delta(F475W-F814W) = 0.40$ mag, and once added to the images, the photometric reduction
followed identical procedures to the steps described above.

The measured CMD for the artificial-star experiments
combining all trials is shown in Figure \ref{fig:fakecmd}. The artificial blue and red
sequences are easily distinguished from one another for any magnitudes $F814W \lesssim 26.5$; fainter than
that, the sequences start to overlap because of the color spread generated purely by measurement
scatter.  As expected for photometry in very uncrowded fields like these, the mean measurement
uncertainties estimated by \emph{allstar} as a function of magnitude agree well with the
estimates from these \emph{addstar} runs.

\begin{figure}[t]
 \vspace{0.0cm}
 \begin{center}
\includegraphics[width=0.5\textwidth]{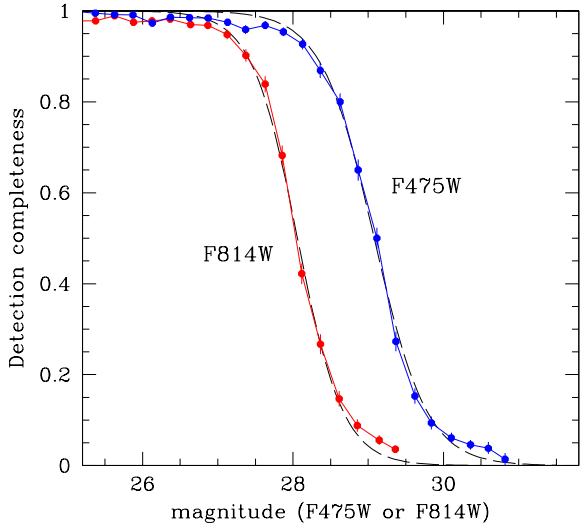}
\end{center}
\vspace{-0.5cm}
\caption{Mean curves for detection completeness $f$ for the radial range $R > 20''$.
	The interpolation function parameters are as defined in Table \ref{tab:f}.
}
\vspace{0.5cm}
\label{fig:f_all}
\end{figure}

The completeness function $f(m)$ is the fraction of inserted stars that were recovered by
the photometry.  The \emph{addstar} 
results for $f$ are shown in Figure \ref{fig:f_all}. 
Once outside the innermost zone
$R > 20''$ (400 px), no significant dependence of $f(m)$ on $R$ is seen; as described in
Paper I, the field is unaffected by crowding at any radii, and the background galaxy
light has already decreased below the point where it affects the completeness.

To describe the shape of the function $f$, it is useful to have
an interpolation curve that follows the data simply and accurately.
A sigmoid-type function of the form
\begin{equation}
	f(m) \, = \, {1 \over {1 + e^{\alpha (m - m_0)} }}
\end{equation}
\noindent satisfies these criteria very well.  Here,
$m_0$ is the magnitude level at which $f = 0.5$
(the 50\% completeness limit) and $\alpha$ is a parameter adjusted
to match the steepness of dropoff of $f(m)$ towards fainter magnitudes. This function is similar
in form to the Fermi/Dirac probability distribution, and is also the same as the formula
used by \citet{alamo-martinez_etal2013} (see their Eq.~2) with C=0.
Other and more complex functional forms can be found, e.g., in
\citet{fleming_etal1995}, \citet{puzia_etal1999}, \citet{barker_etal2004}, and \citet{alamo-martinez_etal2013} 
useful for various special circumstances that fortunately do not apply here.

In Table \ref{tab:f} the detection completeness parameters for the region
$R > 20''$ centered on NGC 6166 (again, excluding only the innermost radial range near galaxy center)
are summarized.

\begin{table*}[t]
\begin{center}
\caption{\sc Detection Completeness Parameters}
\label{tab:f}
\begin{tabular}{llll}
\tableline\tableline\\
\multicolumn{1}{l}{Detector} &
\multicolumn{1}{l}{Filter} &
\multicolumn{1}{l}{$m_0$} &
\multicolumn{1}{l}{$\alpha$} 
\\[2mm] \tableline\\
ACS/WFC & F475W & 29.00 & 2.45 \\
        & F814W & 28.00 & 2.45 \\
WFC3    & F475W & 29.40 & 4.16 \\
        & F814W & 27.50 & 4.30 \\
\\[2mm] \tableline
\end{tabular}
\end{center}
\vspace{0.4cm}
\end{table*}

As noted above, the observed scatter in colors for the real objects in Fig.~\ref{fig:cmd} is much larger
than for the simulation in Fig.~\ref{fig:fakecmd} and is large enough
to obscure any clean division between the standard blue and red GC sequences.  
To test further whether or not the observed scatter is intrinsic, three of the authors
(WEH, JPB, BCW) ran independent photometric reductions in different ways
starting from the raw images.  Comparative tests included different forms of small-aperture photometry 
from DAOPHOT and SE, along with selection criteria that also differed among the three reductions.
All these yielded color-magnitude diagrams that showed close agreement with the \emph{allstar} reductions, to well within 
the internal measurement uncertainties at all magnitudes.
In the following analysis, we therefore continue using the \emph{allstar} data.

Lastly, we emphasize that the analysis presented in the following sections relies
on the magnitude range $F814W < 27$, within which photometric completeness
is high.

\section{The WFC3 Parallel Field}

\begin{figure}[t]
 \vspace{0.0cm}
 \begin{center}
\includegraphics[width=0.5\textwidth]{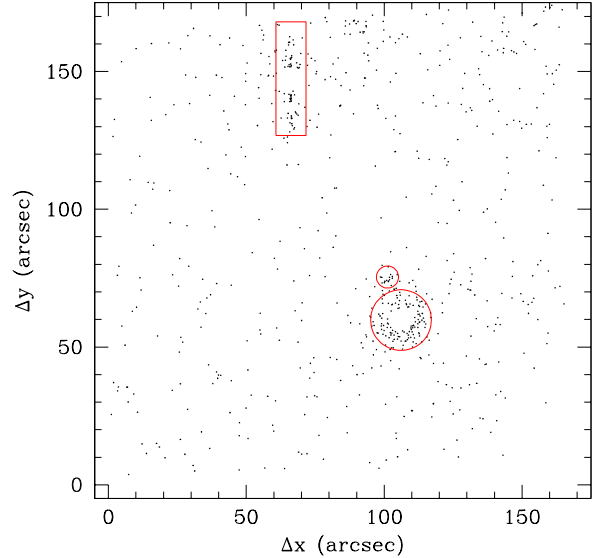}
\end{center}
\vspace{-0.5cm}
\caption{Locations of measured objects in the WFC3 field with 
	$F814W < 27.2$.  Axes are labelled in arcseconds.  The exclusion regions
	around three galaxies in the field are marked as the red circles and box.
}
\vspace{0.5cm}
\label{fig:wfc3_xy}
\end{figure}

The WFC3 camera was used in Coordinated Parallel mode to image a comparison
field centered at ($\alpha$ = 16:28:46.3, $\delta$ = 39:26:46.8) (J2000) through the
same pair of filters.  The projected distance of the WFC3 field center
from NGC 6166 is $6.5'$ SSE, equivalent to 245 kpc, and thus still well
within the volume of the A2199 galaxy cluster.  
Coincidentally -- and very usefully -- the WFC3 field center is at the same radius
as the outermost extent of the surface brightness profile measured
by B15.

\begin{figure}[t]
 \vspace{0.0cm}
 \begin{center}
\includegraphics[width=0.5\textwidth]{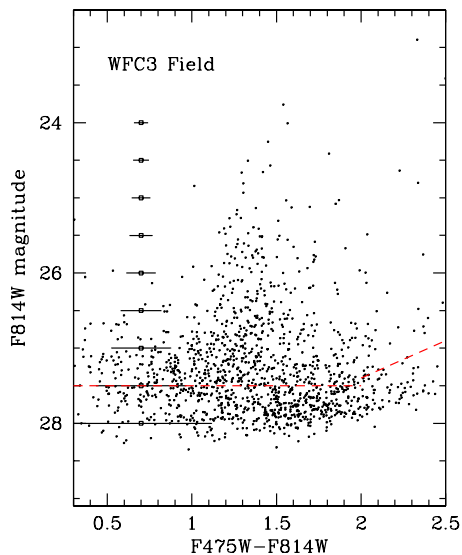}
\end{center}
\vspace{-0.5cm}
\caption{Color-magnitude distribution for the starlike objects in the WFC3 field, lying
	outside the exclusion regions defined in the previous figure.  Internal photometric
	uncertainties for the color indices are indicated along the left side.  The
	red dashed line shows the 50\% photometric completeness limit as determined from
artificial-star tests.}
\vspace{0.5cm}
\label{fig:wfc3_cmd}
\end{figure}

Exactly the same measurement procedure as outlined above was
followed.  The distribution of measured starlike objects brighter than $F814W = 27.2$
is shown in Figure \ref{fig:wfc3_xy}.
Here, excess populations of objects can
be seen grouped closely around three other A2199 galaxies:
these are an edge-on disk galaxy (PGC058278) near the top edge of the frame;
a moderate-sized elliptical (PGC058279/282) at lower right; and a small elliptical
(SDSSJ162845.08+392629.5) just above it.  As shown in the Figure, exclusion regions around
each of these three were drawn and objects within those regions deleted from the lists.
The remaining area outside the exclusion regions is 7.294 arcmin$^2$.

Figure \ref{fig:wfc3_cmd} shows the color-magnitude distribution for the starlike objects 
excluding the ones near these small galaxies.  A population of objects is clearly present in the same range of
colors and magnitudes as the GC blue sequence in Fig.~\ref{fig:cmd}, along with a sprinkling of redder objects.
To answer how many of these are GCs that could belong either to the extended envelope of NGC 6166
or to the A2199 ICM,  we need an estimate of the actual
background density of starlike objects, which is addressed in the next section.

\begin{figure}[t]
 \vspace{0.0cm}
 \begin{center}
\includegraphics[width=0.5\textwidth]{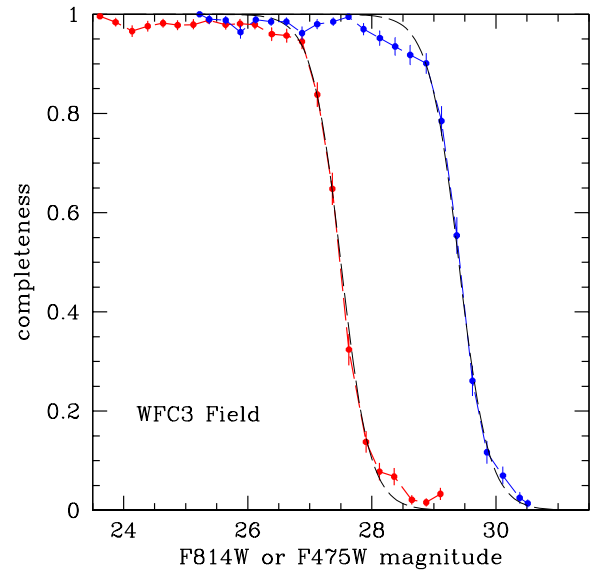}
\end{center}
\vspace{-0.5cm}
\caption{Photometric completeness curves for WFC3 $F814W$ (red points) and
	$F475W$ (blue points).  The dashed lines show the interpolation curves
with parameters $m_0, \alpha$ as listed in Table \ref{tab:f}.}
\vspace{0.5cm}
\label{fig:wfc3_f}
\end{figure}

Artificial-star tests were run in the same way as for the ACS field, to determine the
photometric completeness $f(m)$ and internal measurement uncertainties, with results 
as shown in Figure \ref{fig:wfc3_f}.

\section{Assessment of Field Contamination}

\begin{figure}[t]
 \vspace{0.0cm}
 \begin{center}
\includegraphics[width=0.5\textwidth]{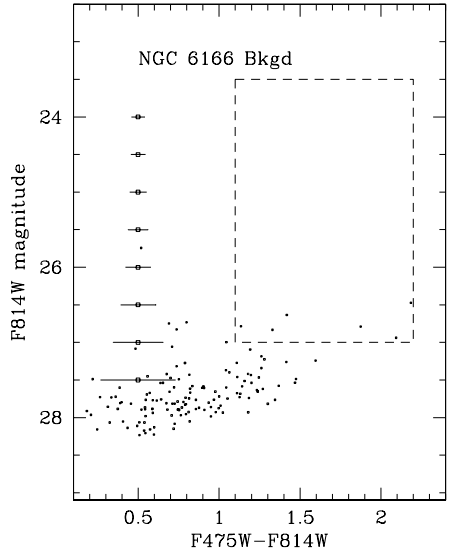}
\end{center}
\vspace{-0.5cm}
\caption{Color-magnitude diagram for a background field near A2199, observed
with the ACS/WFC camera in $F475W,F814W$.  The box outlined by dashed lines
marks the approximate range of magnitudes and colors occupied by globular clusters
around NGC 6166.}
\vspace{0.5cm}
\label{fig:cmd_bkgd}
\end{figure}

Ideally we would like to measure the field contamination level directly, from ACS/WFC or WFC3 images that
(a) use the same filters as in our data, (b) have similar exposure times, and (c) are from pointings close to
the A2199 region in a ``blank field'' not falling on any other cluster of galaxies.
Unfortunately, images matching these criteria are hard to find in the HST Archive
anywhere within several degrees of A2199, but a
field that comes usefully close is from program GO-10412 (PI: Lacy).
From their various ACS pointings around the sky we select their target
at $\alpha =$ 16:56:47.1, $\delta =$ +38:21:36.7, which is at a projected distance
of 5.55 degrees (= 12.6 Mpc) from NGC 6166.
Exposure times are 1876 sec in $F475W$ and 1760 sec in
$F814W$.  
The resulting color-magnitude diagram obtained with the same selection procedures is
shown in Figure \ref{fig:cmd_bkgd}.  Starlike objects are quite rare
particularly for the range of magnitudes ($23.5 \lesssim F814W \lesssim 27.0$)
and color indices ($1.1 < (F475W-F814W) <  2.2$) that generously enclose the
NGC 6166 GC population.  This target box is shown in Figure \ref{fig:cmd_bkgd}.

Aside from the shorter exposure times, a more important difference between this background field and
our data is that its raw exposures used only one quadrant of the ACS/WFC array, and
thus contain only a quarter of the field size we would like.  Nevertheless, after scaling the
areas these results suggest
that no more than $\sim 20$ starlike objects contaminate the CMD
within our main target GC region; and most of these will be fainter than
$F814W \simeq 26.5$. 

\begin{figure}[t]
 \vspace{0.0cm}
 \begin{center}
\includegraphics[width=0.5\textwidth]{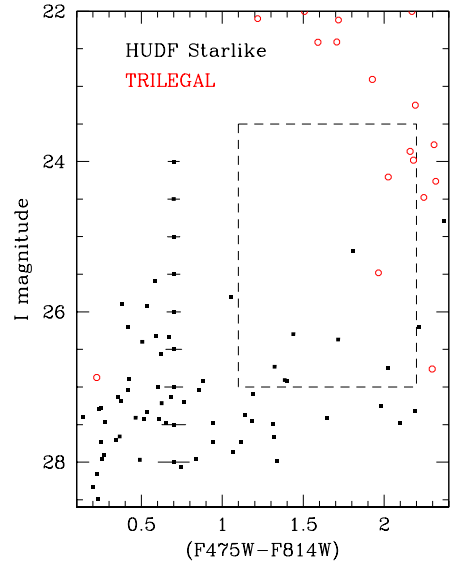}
\end{center}
\vspace{-0.5cm}
\caption{Color-magnitude diagram for starlike objects measured in the HUDF
	(Hubble Ultra-Deep Field) (black points), and for a model population
of foreground stars generated by TRILEGAL (red points; see text). As in the previous
figure, the dashed box indicates the region occupied by NGC 6166 globular clusters.
Errorbars at the left are the measurement uncertainties from the HUDF catalog.}
\vspace{0.5cm}
\label{fig:hudf}
\end{figure}

As a second check, data were used from the Hubble Ultra-Deep Field (HUDF) as provided
in the catalogs at \emph{heasarc.gsfc.nasa.gov/W3Browse/hst/hubbleudf.html}.
The HUDF data in $F435W$ and $F775W$ were transformed to $(B-I)$ with the conversions
in \citet{saha_etal11}, and from there to $(F475W-F814W)$.  To select out only starlike 
objects 
we removed any with $r_{1/2} > 3.0$ px, $ell > 0.3$, or $fwhm > 4.1$ px, leaving just
173 objects over the ACS/WFC field area.  These remaining objects ought to be a combination
of Milky Way foreground stars and faint, very small-scale background galaxies at high redshift.
Their color-magnitude distribution is shown in Figure \ref{fig:hudf}.  Although the color
transformations and selection criteria are not as exact a match as we would like, the results indicate again that
contamination in the target GC region is at the level of less than a dozen objects.

Finally, we have used the TRILEGAL model \citep{girardi_etal2005}
to simulate the expected population of Milky Way foreground stars in the direction of NGC 6166 and over
the ACS/WFC field area.  The results, again transformed into our filter system,
are shown as the red circles in Fig.~\ref{fig:hudf}.  No more than a handful of stars
fall within the target CMD region.

In summary, the contaminating population of foreground stars is mostly brighter and redder
than the NGC 6166 GC population, while the faint, small background galaxies that make it
through our selection criteria are mostly fainter or bluer.  The net field contamination
in the GC region of the CMD  is at the level of $\simeq 0.2$\% and thus
negligible. The overwhelming majority of the objects
in Figs.~\ref{fig:cmd} and \ref{fig:wfc3_cmd} are the globular cluster population we are seeking.
In the following analysis, we do not apply any contamination corrections.
We are now in a position to investigate the spatial and metallicity distributions of the GCS.

\section{Color and Metallicity Distributions}

A physical division of GC populations into distinct metal-poor and metal-rich subgroups was 
first established clearly for the Milky Way \citep{zinn85} by basing the division on
the combination of metallicity, spatial distribution, and kinematics.  Bimodality
was then found in a steadily increasing list       
of other galaxies of every type and environment
\citep[e.g.][among dozens of other papers]{zepf_ashman93,geisler_etal1996,gebhardt_kissler-patig99,larsen_etal01,kundu_whitmore01,peng_etal06,harris09a,brodie_etal2014}
and were linked to various two-phase formation scenarios for major galaxies
\citep[e.g.][]{ashman_zepf92,forbes_etal97,cote_etal98,beasley_etal02,brodie_strader06}.

\begin{figure}[t]
 \vspace{0.0cm}
 \begin{center}
\includegraphics[width=0.5\textwidth]{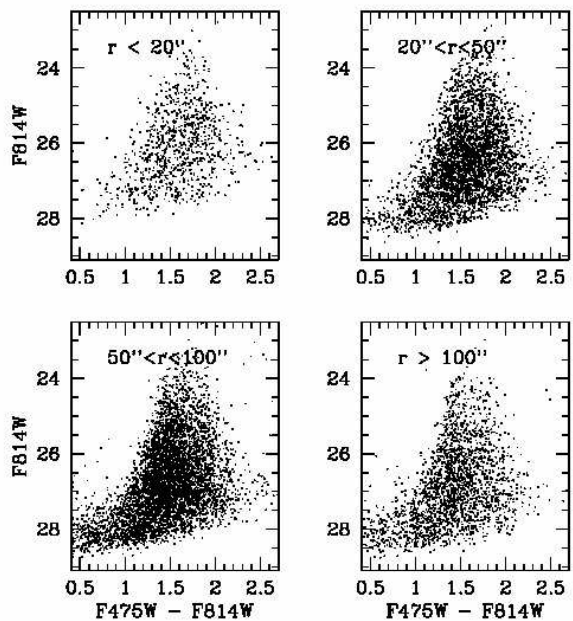}
\end{center}
\vspace{-0.5cm}
\caption{Color-magnitude diagram for the globular cluster population around
NGC 6166, subdivided into four radial zones.}
\vspace{0.5cm}
\label{fig:cmd4}
\end{figure}

A new round of interpretations connects the two subpopulations more directly
to hierarchical-merging galaxy formation models 
\citep{kravtsov_gnedin05,muratov_gnedin10,tonini2013,li_gnedin14}.
Old, metal-poor clusters in a large galaxy 
formed both within the
gas-rich pregalactic dwarfs at the beginning of the merger tree (redshifts $z \gtrsim 4$), and within
dwarf satellite galaxies that were later accreted by the continuously growing
central giant.  By contrast, metal-rich clusters can form
either within more massive halos during the most active epoch of the merger tree ($z \sim 2-3$),
or during later major mergers \emph{if} such mergers bring in significant amounts
of gas (see the references cited above).

\subsection{Bimodality or Not? The Shape of the MDF}

For galaxies well beyond the Local Group, GC metallicity measurements based directly on 
spectroscopic indices are difficult, so in most studies of this type, integrated color indices
are used to measure large samples of GCs more efficiently.  The specific transformation from color to [Fe/H]
depends on the index and in some cases may be measurably nonlinear \citep[e.g.][]{peng_etal06},
but in all cases the color index increases monotonically (becomes redder)  with increasing metallicity.
The \emph{color distribution function} (CDF) is therefore very useful as a proxy for the MDF.

In Figure \ref{fig:cmd4}, the CMDs for four radial zones are shown, now including the
inner region $R < 20''$ and extending to the boundaries of the ACS field.  The bluer
clusters ($(F475W-F814W) \lesssim 1.6$) become relatively more prominent
at larger radii, continuing outward to the WFC3 outer field where they are completely dominant.
Thus with or without bimodality, a radial metallicity gradient is present, as is the case
for most GCSs 
\citep[e.g.][]{geisler_etal1996,rhode_zepf2004,larsen_etal01,harris09a,harris09b,usher_etal2013}.
This result will be discussed below in the details of the spatial distribution.

The more pointed question here is whether
or not we can talk meaningfully about distinct metal-poor and metal-rich components
given the nearly continuous spread of measured colors.   
We apply a two-Gaussian fit to the CDF for the subset
of GCs in the magnitude and color ranges $23.0 < F814W < 26.5$ and $1.1 < (F475W-F814W) < 2.2$
using the GMM code \citep[Gaussian Mixture Modelling; see][]{muratov_gnedin10}.  Since the sample size is large
it is easily possible to carry out the fit
solving for 5 free parameters: the mean colors $\mu_{1,2}$(blue,red), their Gaussian dispersions $\sigma_{1,2}$(blue,red),
and the blue fraction $p_1 \equiv N(blue)/N(tot)$.

\begin{figure}[t]
 \vspace{0.0cm}
 \begin{center}
\includegraphics[width=0.5\textwidth]{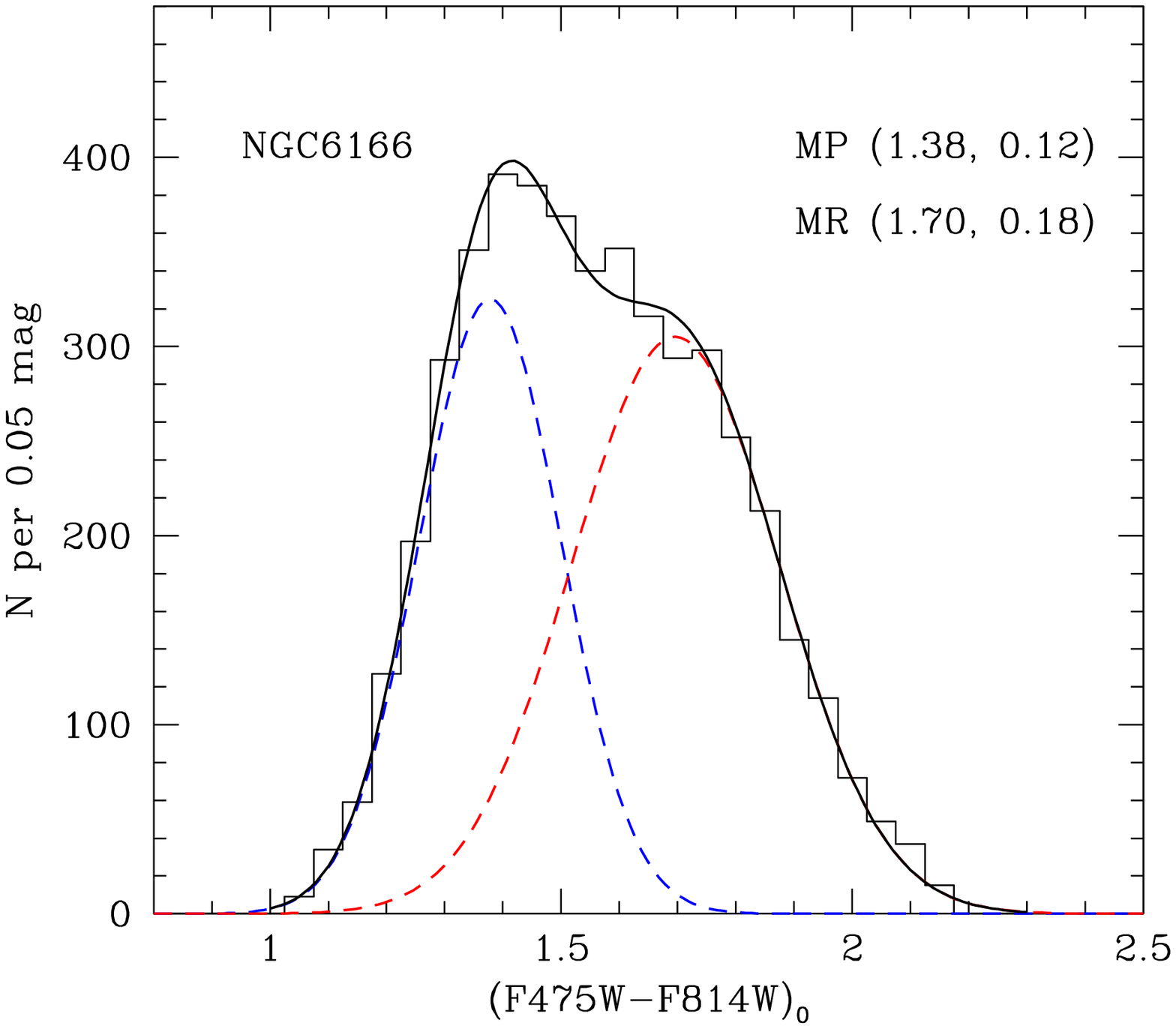}
\end{center}
\vspace{-0.5cm}
\caption{Color histogram for \emph{dereddened} color indices of the NGC 6166 globular clusters, in bins
of 0.05 mag.  All clusters in the ACS field brighter than $F814W = 26.5$ at all radii are included.
A best-fit bimodal Gaussian model is superimposed, where the metal-poor component
is shown as the blue dashed line, the metal-rich component as the red dashed line, and their sum
as the solid line.  The mean and dispersion of the two components are given at upper right.}
\vspace{0.5cm}
\label{fig:chisto}
\end{figure}

The results, plotted on the color histogram for all radii $R$, are shown in Figure \ref{fig:chisto}.
The best-fit solution yields ($\mu_1 = 1.401 \pm 0.010$, $\mu_2 = 1.715 \pm 0.015$) 
for the means, ($\sigma_1 = 0.121 \pm 0.005$, 
$\sigma_2 = 0.178 \pm 0.006$) for the dispersions, and $p_1 = 0.422 \pm 0.038$ for the blue fraction. 
Uncertainties in the fitted quantities were determined through bootstrapping.
(Note that the values plotted in Fig.~\ref{fig:chisto} are the \emph{dereddened} color indices,
not the raw colors.)

The bimodal Gaussian model turns out to provide an extremely close match to the data, even though the
two modes are heavily overlapped.  A single-Gaussian
fit is strongly rejected by GMM at far above 99\% significance. In addition, a \emph{trimodal} 
model provides no improvement to the
total fit, and in any case the third mode identified by GMM turns out to be on the red-side
tail and makes up only 4\% of the total population.

\begin{figure}[t]
 \vspace{0.0cm}
 \begin{center}
\includegraphics[width=0.5\textwidth]{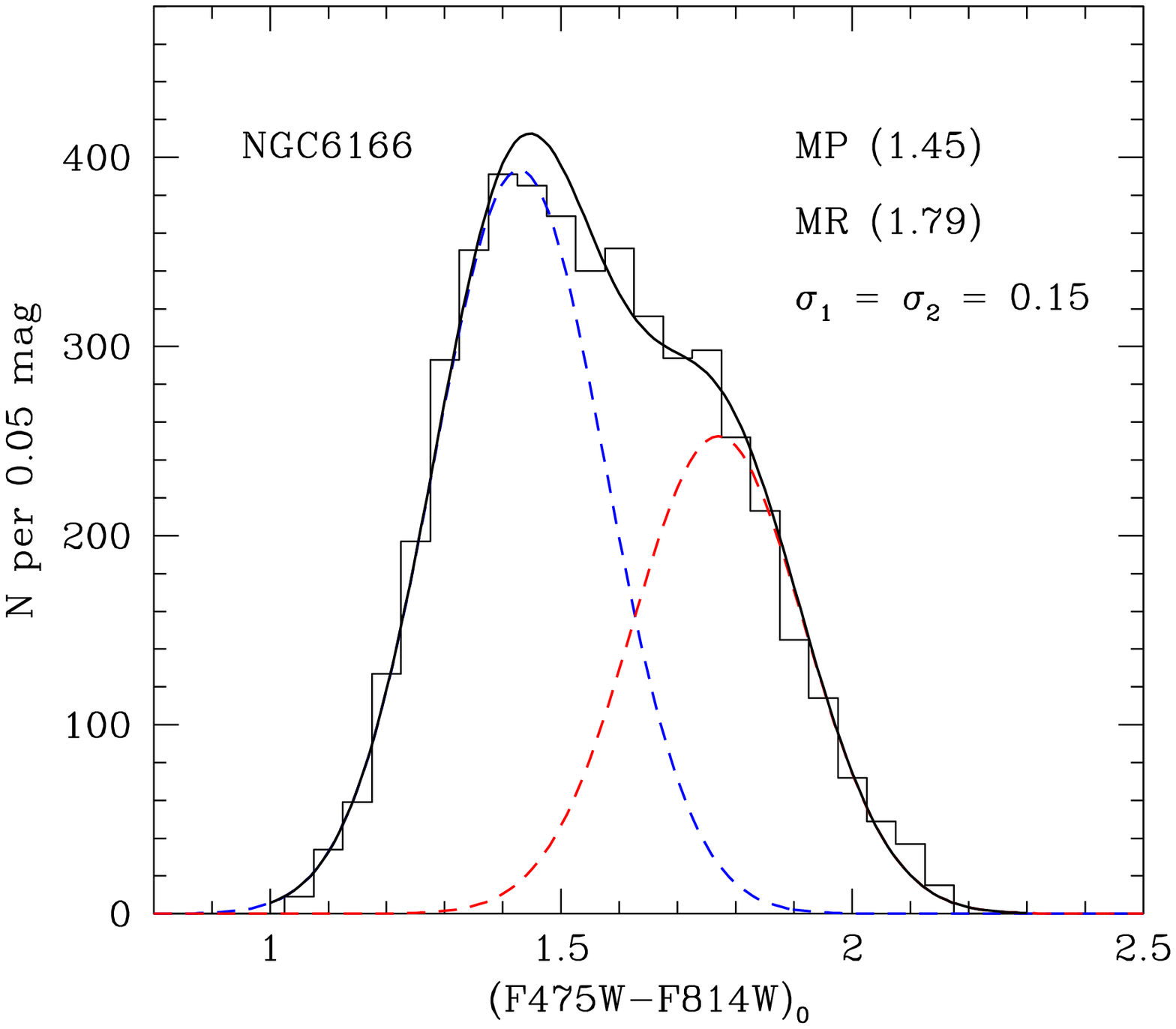}
\end{center}
\vspace{-0.5cm}
\caption{Color distribution for the NGC 6166 GCs, with a bimodal Gaussian
where both components are required to have the same variance (homoscedastic).}
\vspace{0.5cm}
\label{fig:chisto_homo}
\end{figure}

If we enforce a homoscedastic fit (same variances for both modes), as has frequently been
done in the previous literature, the GMM solution yields
$\mu_1 = 1.451$, $\mu_2 = 1.793$, $\sigma_1 \equiv \sigma_2 = 0.147$, $p_1 = 0.609$.
This result is shown in Figure \ref{fig:chisto_homo}.  This second solution 
does not do as well especially at matching the blue peak and intermediate color range, and
the heteroscedastic model (different variances) is strongly preferred by GMM 
at a $>$99.9\% level of significance.
In addition, we have no \emph{a priori} physical reason to impose equal variances.  As noted by \cite{peng_etal06},
the forced assumption $\sigma_1 \equiv \sigma_2$ often has the unwanted consequence
of driving the estimated peaks $\mu_{1,2}$ away from their true values.  
The most striking difference between the two fits
is the surprisingly large change in the relative numbers of clusters in each mode:  imposing
equal variances increases the ratio $p_1$ from 0.42 to 0.61.  A difference this large
was numerically possible essentially because of the 
overlap between the two modes.

As a check of the numerics we have also used the multimodal fitting code
RMIX \citep[see][]{wehner_etal08,harris09a}, which yielded 
results completely consistent with GMM.
RMIX permits the use of any number of modes, as well as other non-Gaussian asymmetric functions such as Poisson
or gamma functions, but none of these yield improvements on the bimodal Gaussian model.

\begin{table*}[t]
\begin{center}
\caption{\sc Bimodal Gaussian Fits by Radial Range}
\label{tab:gmm_rad}
\begin{tabular}{lrlllll}
\tableline\tableline\\
\multicolumn{1}{l}{$R$ Range} &
\multicolumn{1}{r}{$n$} &
\multicolumn{1}{l}{$\mu_1 (\pm)$} &
\multicolumn{1}{l}{$\mu_2 (\pm)$} &
\multicolumn{1}{l}{$\sigma_1 (\pm)$} &
\multicolumn{1}{l}{$\sigma_2 (\pm)$} &
\multicolumn{1}{l}{$p_1 (\pm)$} 
\\[2mm] \tableline\\
$< 20''$ & 558 & 1.386(0.055) & 1.755(0.038) & 0.129(0.023) & 0.178(0.016) & 0.29(0.12) \\
$20''-40''$ & 1138 & 1.442(0.036) & 1.757(0.128) & 0.128(0.014) & 0.178(0.015) & 0.35(0.11) \\
$40''-70''$ & 1509 & 1.457(0.021) & 1.763(0.036) & 0.133(0.009) & 0.176(0.013) & 0.48(0.09) \\
$70''-100''$ & 945 & 1.419(0.016) & 1.741(0.035) & 0.114(0.010) & 0.168(0.015) & 0.52(0.07) \\
$100''-165''$ & 611 & 1.400(0.013) & 1.705(0.031) & 0.111(0.007) & 0.170(0.015) & 0.55(0.07) \\
\\
$390''$ (WFC3) & 147 & 1.324(0.021) & 1.674(0.074) & 0.136(0.033) & 0.244(0.013) & 0.71(0.12) \\

\\[2mm] \tableline
\end{tabular}
\end{center}
\vspace{0.4cm}
\end{table*}

\begin{table*}[t]
\begin{center}
\caption{\sc Bimodal Gaussian Fits by Magnitude Interval}
\label{tab:gmm_mag}
\begin{tabular}{lrlllll}
\tableline\tableline\\
\multicolumn{1}{l}{$F814W$ Range} &
\multicolumn{1}{r}{$n$} &
\multicolumn{1}{l}{$\mu_1 (\pm)$} &
\multicolumn{1}{l}{$\mu_2 (\pm)$} &
\multicolumn{1}{l}{$\sigma_1 (\pm)$} &
\multicolumn{1}{l}{$\sigma_2 (\pm)$} &
\multicolumn{1}{l}{$p_1 (\pm)$} 
\\[2mm] \tableline\\
23.0-24.0 & 139 &             &  1.679(0.013) &             &  0.141(0.011)&  0.00 \\
24.0-24.5 & 257 & 1.445(0.011) & 1.687(0.020) & 0.039(0.011) & 0.158(0.011) & 0.22(0.06) \\
24.5-25.0 & 531 & 1.431(0.021) & 1.680(0.032) & 0.066(0.018) & 0.173(0.014) & 0.22(0.10) \\
25.0-25.5 & 848 & 1.456(0.019) & 1.784(0.029) & 0.112(0.009) & 0.158(0.014) & 0.46(0.07) \\
25.5-26.0 & 1255 &1.426(0.016)&  1.781(0.026)&  0.122(0.008)&  0.167(0.012)&  0.50(0.06) \\
26.0-26.5 & 1738 &1.380(0.017)&  1.708(0.025)&  0.119(0.009)&  0.206(0.008)&  0.37(0.06) \\
26.5-27.0 & 2044 &1.348(0.017)&  1.708(0.023)&  0.125(0.008)&  0.220(0.008)&  0.41(0.05) \\
\\
23.0-26.5 (PSF) & 4712 & 1.401(0.009) &  1.719(0.015) &  0.122(0.005) &  0.178(0.006) &  0.42(0.04) \\
23.0-26.5 (ap) & 3696 & 1.464(0.011) &  1.783(0.017) &  0.123(0.006) &  0.163(0.008) &  0.48(0.04) \\

\\[2mm] \tableline
\end{tabular}
\end{center}
\vspace{0.4cm}
\end{table*}

\subsection{Color (Metallicity) versus Radius and Luminosity}

\begin{figure}[t]
 \vspace{0.0cm}
 \begin{center}
\includegraphics[width=0.5\textwidth]{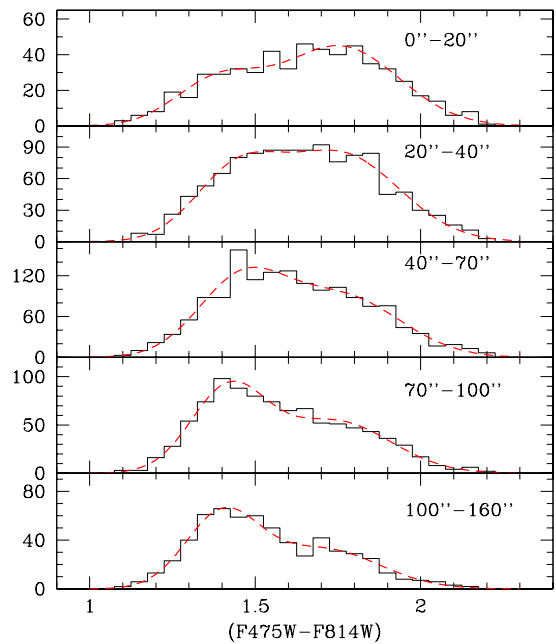}
\end{center}
\vspace{-0.5cm}
\caption{Color histograms for radial zones centered on NGC 6166.  In each panel
the \emph{dashed red line} gives the best-fit bimodal Gaussian solution from GMM, with
parameters as given in Table \ref{tab:gmm_rad}.  A strong radial metallicity
gradient is present, driven by the changing ratio of blue to red GCs with $R$.}
\vspace{0.5cm}
\label{fig:chisto_rad}
\end{figure}

\begin{figure}[t]
 \vspace{0.0cm}
 \begin{center}
\includegraphics[width=0.5\textwidth]{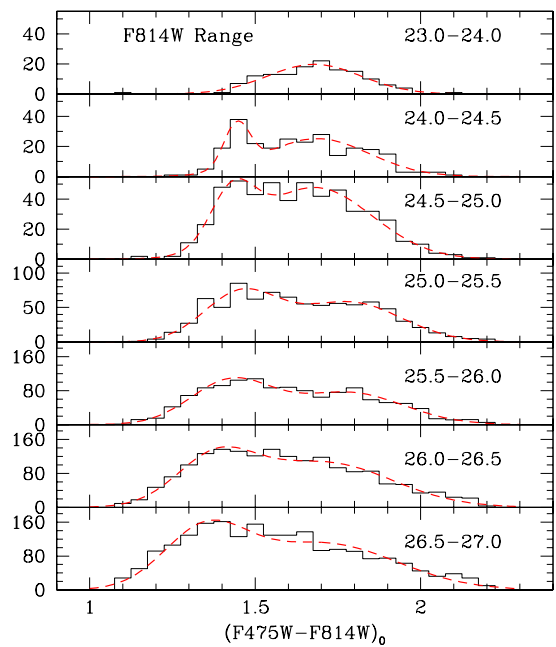}
\end{center}
\vspace{-0.5cm}
\caption{Color histograms for GCs over all radii but subdivided by magnitude.
	The \emph{dashed red line} in each panel shows the best-fit bimodal Gaussian
solution from GMM, with parameters given in Table \ref{tab:gmm_mag}.
Note the shift towards a unimodal distribution in the brightest bin.}
\vspace{0.5cm}
\label{fig:chisto_mag}
\end{figure}

GMM fitting results for samples subdivided by radial zone $R$, and by
magnitude range $F814W$, are listed in Tables \ref{tab:gmm_rad} and \ref{tab:gmm_mag}, and the full
histograms are displayed in Figures \ref{fig:chisto_rad} and \ref{fig:chisto_mag}.  
For the radial zones in Table \ref{tab:gmm_rad}, the data in the range $23.0 < F814W < 26.5$ and $1.1 < (F475W-F814W) < 2.2$ are
used.  Successive columns give (1) the zone boundaries in arcseconds; (2) the total number
of GCs in the range; (3,4) the means and uncertainties
of the blue and red modes; 
(5,6) the standard deviations of the blue and red modes and their uncertainties; and
(7) the blue GC fraction $p_1$.  

The last line in Table \ref{tab:gmm_rad} gives the parameters for the
WFC3 field:  here, two modes are still present and both modes are nominally slightly bluer than in the ACS radial bins.
However, we believe it is risky to conclude that these differences are intrinsic, given 
the remaining uncertainties in the filter zeropoints of both cameras, and the internal zeropoint
corrections of the PSF-based photometry to large apertures.

The last two lines in Table \ref{tab:gmm_mag} give the GMM fit parameters derived separately from
the \emph{allstar} PSF-fitting photometry, and then the small-aperture photometry described above.
The zero-point calibrations for the aperture photometry were only approximate and thus account
for the differences in $\mu_1,\mu_2$; more importantly, the internal dispersions $\sigma_1, \sigma_2$ 
are nearly identical from both methods, indicating again that the observed color spread is not an artifact 
of the photometric method.

\begin{figure}[t]
 \vspace{0.0cm}
 \begin{center}
\includegraphics[width=0.5\textwidth]{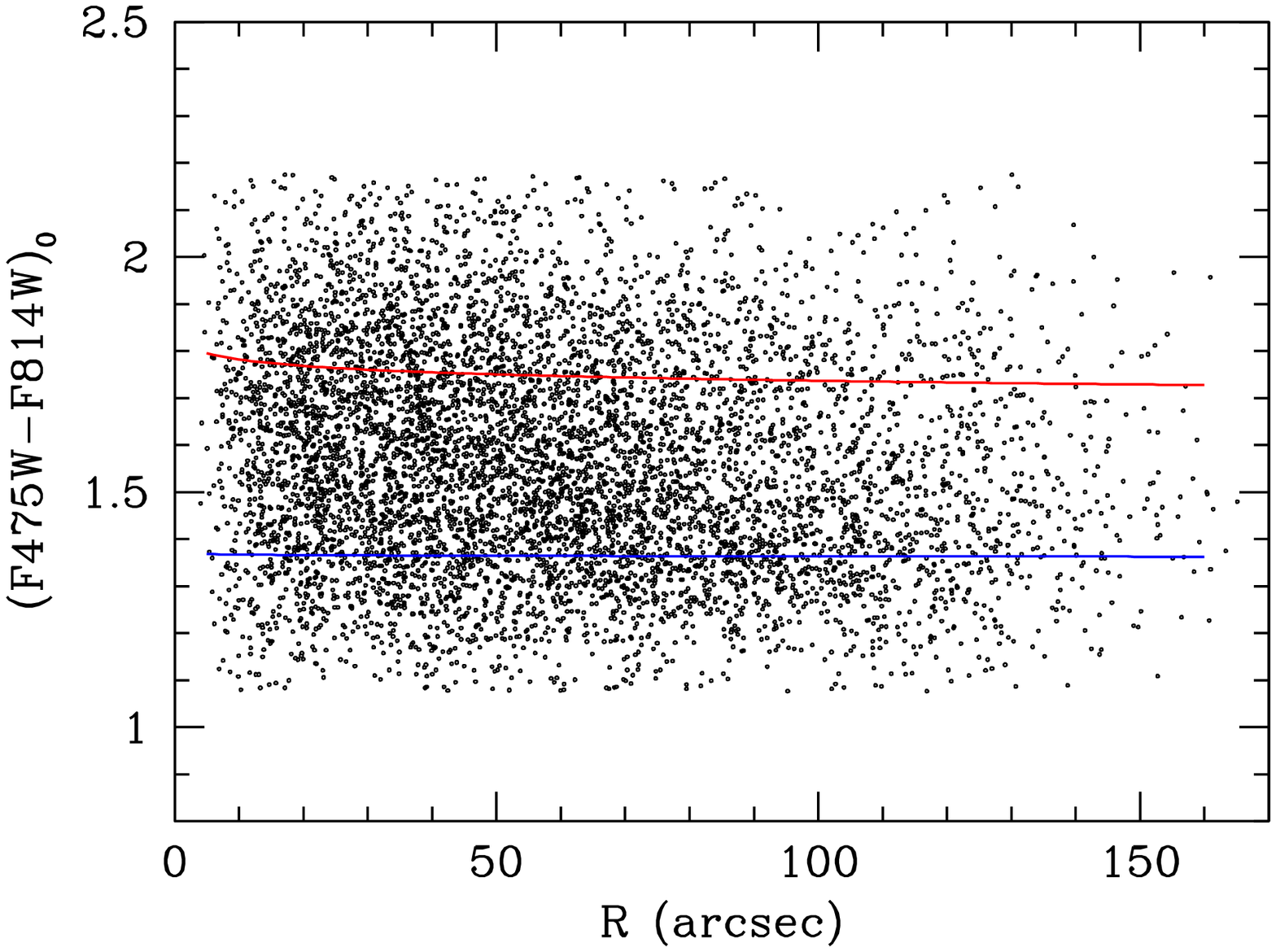}
\end{center}
\vspace{-0.5cm}
\caption{Dereddened color index versus radius for the GCs in NGC 6166.  The blue and
red lines show the solutions for color versus radius as listed in the text.}
\vspace{0.5cm}
\label{fig:radcolor}
\end{figure}

Finally, in Figure \ref{fig:radcolor} we show the GC intrinsic color versus radius $R$, for all GCs brighter
than $I = 27$.  Dividing the sample into two at $(F475W-F814W) = 1.55$ (see below), and then carrying out
a solution for mean color versus log $R$, yields 
$\langle F475W-F814W \rangle_0 (blue) \, = \, (1.37 \pm 0.01) - (0.003 \pm 0.007) log R''$ and
$\langle F475W-F814W \rangle_0 (red) \, = \, (1.83 \pm 0.02) - (0.045 \pm 0.009) log R''$.
The mean color of the MP population shows no significant change with radius,
while the MR population exhibits a shallow but significant negative gradient equivalent
to a heavy-element abundance gradient $Z \sim R^{-0.14 \pm 0.03}$.
By contrast, the GCSs of many other giant galaxies show negative metallicity
gradients of similar amplitudes in both their 
MP and MR components \citep{geisler_etal1996,forte_etal2001,lee_etal2008,harris09a,harris09b,hargis_rhode2014}.
For NGC 6166, \emph{the overall metallicity gradient in the global GCS is generated primarily by
a ``population gradient'' of the changing ratio of blue to red GCs.}

The absence of a metallicity gradient for the MP clusters is suggestive of strong mixing of the cluster sub-populations          
brought by former satellite galaxies accreted by the central host.  
Recently \citet{monachesi_etal2015} have found that Milky-Way-type disk galaxies display a wide range of
halo metallicity gradients out to remarkably large radius ($\sim 50$ kpc or more than 10 effective radii).
They find that large galaxy-to-galaxy differences also exist in the width (intrinsic dispersion) of the
halo-star MDFs.  For giant galaxies like the BCGs that may have accreted many disk galaxies or stripped stars
via harassment, this material is consistent with the idea that their halos would have ended up with strongly mixed stellar
populations and weak global metallicity gradients.

\begin{figure}[t]
 \vspace{0.0cm}
 \begin{center}
\includegraphics[width=0.5\textwidth]{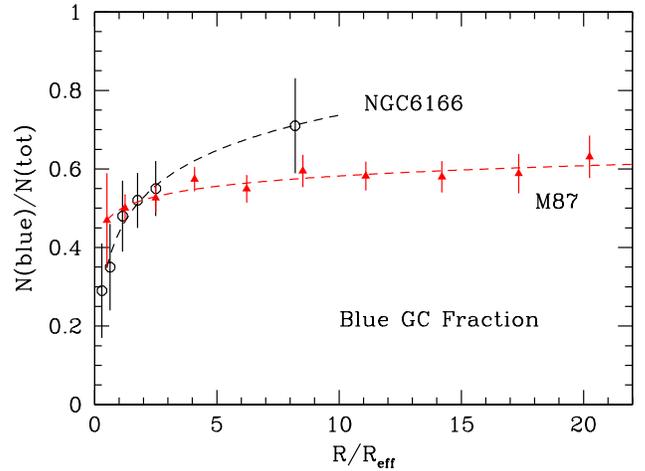}
\end{center}
\vspace{-0.5cm}
\caption{Increase of the metal-poor GC fraction (NGC 6166, open circles and errorbars) with projected radius $R$.
The equation for the fitted profile (dashed line) is given in the text.  Solid diamonds and red dashed line give
the same trend for the M87 GCS.  Note that the radius is labelled in units of $R_{eff}$, the effective radius
of the halo light profile.
}
\vspace{0.5cm}
\label{fig:fblue}
\end{figure}

The change in blue fraction $p_1$ with $R$ turns out empirically to behave in an extremely
simple way.  The trend is shown in Figure \ref{fig:fblue}, with datapoints taken from
Table \ref{tab:gmm_rad}.  Numerically we find an excellent fit to a logarithmic form,
\begin{equation}
	p_1(N6166) \, \equiv \, N(blue)/N(tot) \, = \, 0.30 \, {\rm log} (29.2 R / R_{eff}) 
\end{equation}
where $R_{eff} = 0.792' = 30$ kpc is the effective radius of the halo light profile (B15).
This curve is shown in Figure \ref{fig:fblue}, and will be used below to help derive
the radial profiles of the MP and MR subsystems.

\subsection{Comparison with M87 and Other BCGs}

\begin{figure}[t]
 \vspace{0.0cm}
 \begin{center}
\includegraphics[width=0.5\textwidth]{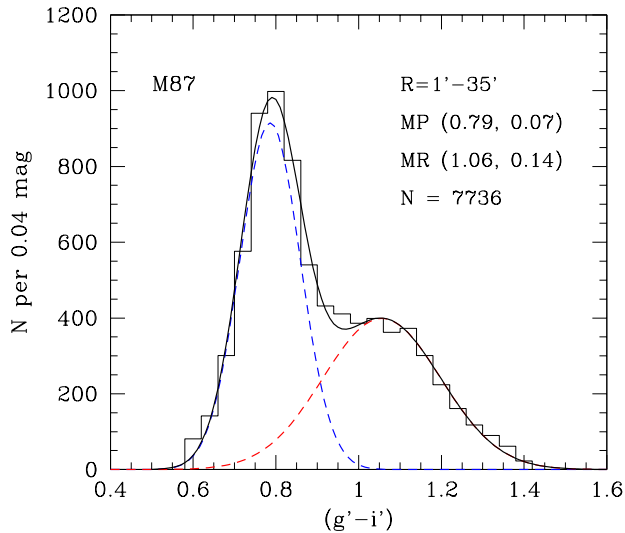}
\end{center}
\vspace{-0.5cm}
\caption{Color distribution function for the GCs in M87, with data from
	\citet{harris09b}.  The bimodal Gaussian fit given in the text is shown in the dashed lines,
with values for the mode peaks and dispersions listed at upper right.}
\vspace{0.5cm}
\label{fig:gmm_m87}
\end{figure}

Part of the interest in this study is the large spatial coverage of the GCS, which
allows us to detect large-scale radial trends that may be difficult to see in nearer or smaller galaxies.
Though similarly deep, wide-field studies are still unusual, a particularly good comparison case is M87, the Virgo BCG.  
We use the study of M87 carried out in
the $g',i'$ bands from \citet{harris09b} to analyze its MDF as a function of $R$ in the same way as for
NGC 6166.  To minimize field contamination we select the M87 data in the range $20 < i' < 23$ and
$0.6 < (g'-i') < 1.4$.  The color histogram for 7736 total objects in this range and over projected distances
$1' < R < 35'$ (4 to 160 kpc) is shown in Figure \ref{fig:gmm_m87}.  The bimodal-Gaussian GMM best-fit solution provides
an excellent and well determined match to the color-index distribution, with
mode peaks ($\mu_1 = 0.786 \pm 0.002, \mu_2 = 1.056 \pm 0.007$), dispersions ($\sigma_1 = 0.073 \pm 0.002, \sigma_2 = 0.142 \pm 0.003$),
and blue fraction $p_1 = 0.541 \pm 0.015$.  

The two modes in M87 are more distinct and better separated than for NGC 6166.  An objective
measure of the separation is the $D-$statistic \citep{muratov_gnedin10},
\begin{equation}
	D \, = \, {{(\mu_2 - \mu_1)} \over {[(\sigma_1^2 + \sigma_2^2)/2]^{1/2} }} \, .
\end{equation}
For M87, $D = 2.39 \pm 0.08$ whereas for NGC 6166, $D = 2.08 \pm 0.12$.  Both are above the $D \simeq 2$
threshold that indicates intrinsic bimodality, but NGC 6166 has relatively broader MP and MR components
that cause stronger overlap.

Bimodal fits to the M87 distribution were done for several radial zones, with the resulting 
trend for $p_1$ as shown in Fig.~\ref{fig:fblue}.
As for NGC 6166, the smooth increase in $p_1$ with $R$ matches a logarithmic form quite well:
\begin{equation}
	p_1(M87) \, = \, 0.087 \, {\rm log} (4.9 \times 10^5~ R / R_{eff}) \, 
\end{equation}
where $R_{eff}(M87) = 1.58' = 7.4$ kpc.  In units of $R_{eff}$, we can trace $p_1$ out further for M87,  
but its outward increase is distinctly shallower. 
The Virgo cluster is dynamically younger than A2199, still actively accreting galaxies, and thus its 
central galaxy may have accreted fewer small satellites that would preferentially add metal-poor clusters to
the M87 outer halo.

Valuable, though less detailed, comparisons of the MDF with those in other galaxies can 
be made after conversion from color to metallicity [Fe/H].
Various color indices have been employed in the recent literature; translations
of any one of them to [Fe/H] are discussed in many papers and, in general,
are not yet as accurate as we would like them to be
\citep[see][for illustrative discussions partly based on modelling]{blakeslee_etal2010,fensch_etal2014,li_gnedin14,vanderbeke_etal2014}.
In some cases, particularly the $(g-z)$ index used for the Virgo and Fornax GCS surveys, 
the transformation to [Fe/H] has noticeable nonlinearity \citep{peng_etal06,blakeslee_etal2010,vanderbeke_etal2014,li_gnedin14}.
For our index $(B-I)$, 
recent linear conversions to [Fe/H] are given by \citet{barmby_etal2000} and \citet{harris_etal06}
calibrated from Milky Way and M31 clusters, and are in close agreement.  
Within the scatter of these calibrations no nonlinearity is evident.  
To be consistent with the analysis of previous BCGs from \citet{harris09a} we use the one
in \citet{harris_etal06}, which transforms to
\begin{equation}
	(F475W-F814W)_0 \, = \, 0.316 {\rm [Fe/H]} + 1.822 \, .
\end{equation}
The conversion is calibrated from 95 Milky Way GCs with well known $(B-I)$ colors, reddenings,
and spectroscopic metallicities.

The converted bimodal MDF parameters are listed in Table \ref{tab:mdf_bcg},
along with eight other BCGs drawn from previous papers.
The successive columns are (1,2,3) galaxy name, host galaxy cluster, and luminosity,
(4) the blue fraction $p_1$,
(5,6) the dispersions for the MP and MR modes, 
(7) the difference $\Delta$[Fe/H] between the mean metallicities
of the MP and MR modes, (8) mode separation $D$, and (9) literature source.
The last line in the Table gives the mean values for the various quantities,
\emph{not including NGC 6166}.

The other galaxies listed are the centrally dominant objects in
clusters of galaxies of various richnesses, from relatively nearby
groups (NGC 1407) to larger systems such as Virgo and Centaurus (M87, NGC 4696).  
Although other discussions of the GCSs for some of these galaxies
are available \citep{forbes_etal2006,mieske_etal2006,mieske2010,bassino_etal2008,peng_etal2009,fensch_etal2014},
the sources used here treat the MDF fits with the same methodology as our NGC 6166 analysis,
and the photometry for these other galaxies has very similar internal uncertainties at the same \emph{absolute} 
magnitude level in the CMD.

\begin{table*}[t]
\begin{center}
\caption{\sc MDF Parameters for BCGs}
\label{tab:mdf_bcg}
\begin{tabular}{lllllllll}
\tableline\tableline\\
\multicolumn{1}{l}{Galaxy} &
\multicolumn{1}{l}{Host Cluster} &
\multicolumn{1}{l}{$M_V^T$} &
\multicolumn{1}{l}{$p_1$} &
\multicolumn{1}{l}{$\sigma_1$(dex)} &
\multicolumn{1}{l}{$\sigma_2$(dex)} &
\multicolumn{1}{l}{$\Delta$[Fe/H]}  &
\multicolumn{1}{l}{$D$}  &
\multicolumn{1}{l}{Source} 
\\[2mm] \tableline\\
NGC 6166 & A2199 & $-23.7$  & 0.57 &  0.39  & 0.56 & 1.01 & 2.08 & 1 \\
\\
M87      & Virgo & $-22.42$ & 0.54 & 0.24   & 0.29 & 0.91 & 2.39 & 2 \\
NGC 1399 & Fornax & $-22.02$ & 0.63 & 0.23 & 0.36 & 0.94 & 3.11 & 3 \\
NGC 1407 & Eridanus & $-22.35$ & 0.33 & 0.28 &  0.39 & 1.18 & 3.10 & 4 \\
NGC 3348 & CfA 69 & $-22.13$ & 0.49 & 0.20 & 0.42 & 1.09 & 2.95 & 4 \\
NGC 3258 & Antlia & $-21.87$ & 0.52 & 0.23  &  0.38 & 0.98 & 3.08 & 4 \\
NGC 3268 & Antlia & $-21.96$ & 0.48 & 0.19  &  0.45 & 0.93 & 3.03 & 4 \\
NGC 4696 & Cen30 & $-23.31$ & 0.51 & 0.26  &  0.45 & 1.00 & 2.87 & 4 \\
NGC 7626 & Pegasus I & $-22.35$ & 0.35 & 0.28  & 0.42 & 1.13 & 2.86 & 4 \\
\\
Mean     &           & $-22.30$ & 0.48 & 0.24 & 0.40 & 1.02 \\
\\[2mm] \tableline
\end{tabular}
\tablecomments{Sources:  (1) This paper (2) Harris 2009b (3) \citet{kim_etal2013} (4) Harris 2009a.
The last line gives the mean for the 8 BCGs excluding NGC 6166.}
\end{center}
\vspace{0.4cm}
\end{table*}

Small differences in photometric zeropoint calibrations for the various color indices 
used in these different studies, plus subsequent transformation into metallicity,
make it difficult to compare the absolute [Fe/H](MP,MR) values meaningfully 
\citep[see also][for discussion of intrinsic galaxy-to-galaxy scatter in the correlations between color and spectral indices]{usher_etal2015}. The more
robust results are the ones in the Table, i.e. the dispersions $\sigma(MP,MR)$ and the offsets
$\Delta$[Fe/H] between the two modes, along with the resulting $D-$statistic.
NGC 6166 is the most luminous BCG in the
list, and it has the broadest metallicity dispersions
for both MP and MR components of all galaxies in the list.
However, the mean metallicity difference between
the MP and MR modes is virtually identical for all of them, at
$\langle \Delta$[Fe/H]$\rangle = 1.0$ dex.

The very high intrinsic dispersions that we see in the NGC 6166 MDF, and the remarkably consistent 1.0 dex     
offset between the two modes, find some theoretical motivation particularly in the recent models 
of \citet{li_gnedin14}.
In these models, GC formation is assumed to be driven by mergers between gas-rich galaxies.          
Clusters inherit the metallicity of their parent galaxy at the moment of formation, 
which is calculated via the observed relation between          
galaxy stellar mass and mean metallicity.  The bulk of the metal-poor clusters come from almost continuous, early     
mergers among small halos at the epochs when they are extremely gas-rich.  To create a GC that is         
massive enough to survive dynamical disruption until the present, the galaxy mass 
needs to be above a certain threshold, $\gtrsim 10^9 M_\sun$, which in turn sets the minimum metallicity.  
The red GCs are contributed by more massive galaxies, in which
the metallicity scales weakly with mass.  Thus, the mean metallicities of the MP and MR modes 
increase only slightly with galaxy mass, maintaining a $\sim 1$ dex offset close to what is observed.  
In contrast, the dispersions of both modes increase with galaxy mass, as the numbers 
of contributing mergers increase. 
This can be seen in Figures 6 and 13 in \citet{li_gnedin14}.
For the giant, central cluster galaxies, both MP and MR modes are so wide that they 
form one broad distribution, reminiscent of what we see for NGC 6166.                                                         

\subsection{Mass-Metallicity Relations}

A more recently discovered feature of interest is the 
trend of mean metallicity with GC luminosity. 
In several large galaxies, along the blue MP sequence especially, 
the mean metallicity has been observed to increase gradually with GC mass
\cite[][among others]{harris_etal06,strader_etal06,mieske_etal2006,wehner_etal08,harris09a,peng_etal2009,cockcroft_etal2009,fensch_etal2014}.
This \emph{mass-metallicity relation} (MMR) is subtle enough that it is still unclear 
if the effect is a simple power law (that is, if we can write $Z \sim M^{\gamma}$
for heavy-element abundance $Z$ with $\gamma = const$), or if the index $\gamma$ itself increases with mass
such that the MP sequence curves more strongly toward the MR sequence at progressively higher 
mass.\footnote{By definition,
$\gamma = \Delta{\rm log}Z / \Delta{\rm log} M$.  In the observational plane, if $\Delta(color) = \alpha \Delta{\rm [Fe/H]}$
and $\Delta(color) = \beta \Delta(mag)$, then $\gamma = -2.5  \beta / \alpha$. This basic version of the transformation
assumes no significant change in the GC mass-to-light ratio with cluster mass.}

The amplitude of the effect may also differ from one galaxy to another, and may be virtually absent in some cases,
notably NGC 4472 and NGC 1399 \citep{strader_etal06,mieske_etal2006,forte_etal2007}.
In addition, for some galaxies (like M87 or NGC 4696) the MP sequence shows a steady near-linear slope in color
toward the red for the brightest $2-3$ magnitudes of the GCLF, but in others such as NGC 1399 or NGC 4696,    
above a certain threshold luminosity near $2 \times 10^6 L_{\odot}$,
the total CDF becomes broad and unimodal rather than bimodal \citep[e.g.][]{dirsch_etal2003,bassino2006,harris09a}.

Empirically, the effect is most noticeable for BCGs where the GCLF is rich enough to populate the highest mass  
range $M > 10^6 M_{\odot}$ thoroughly.  For that reason, in smaller galaxies any MMR is extremely difficult
to identify.  For example, in the Milky Way the only individual GC that
lies clearly in this high-mass regime is $\omega$ Centauri, an object that is well known to contain a
complex set of stellar subpopulations more extreme than in any lower-mass GC \citep[e.g.][]{bellini_etal2009}.

Theory addressing the MMR phenomenon is still in early stages.
A model based on internal GC self-enrichment \citep{bailin_harris09} is capable of matching 
\emph{some} of the various forms taken by the MMR
\citep[see][]{mieske_etal2006,mieske_etal2010,fensch_etal2014}.  One of the strongest motivations for
pursuing models involving some form of extra enrichment that increases with GC mass is that the 
MMR is visible along the blue GC sequence but not the red sequence.
If some extra heavy elements are added to a GC in amounts
depending only on its mass, the visible effect on the integrated colors can be quite noticeable
for a GC that was originally very metal-poor, but nearly negligible for one that was already
metal-rich.
However, because it is driven by local conditions inside the GC during its formation, 
the Bailin/Harris model has difficulty reproducing 
the wide range of MMR slopes seen in different galaxies, which suggests that some environmental
feature must also be at work.
Large galaxies with no MMR slope, and those with broad unimodal MDFs at high GC mass, also remain challenging for this
type of model.  

\citet{vandalfsen_harris2004} and \citet{forte_etal2007} adopt a simpler numerical approach 
to model the MDF that invokes what is essentially
\emph{pre-}enrichment.  They model the MDF of each of the blue and red GCs as 
$dn/dZ \sim exp[-(Z-Z_i)/Z_s]$ for the distribution of heavy-element abundance $Z$,
where the free parameters are the initial or minimum allowed GC metallicity $Z_i$ and
a scale $Z_s$.  This form is analogous to the Simple Model of chemical evolution where
$Z_s$ represents an effective yield.  VanDalfsen \& Harris assumed that $Z_i$ for each
sequence does not change with GC mass. But Forte et al.~note that if
$Z_i$ is a function of GC mass, i.e. if more massive clusters are formed
from more enriched gas, then a MMR with a range of observed forms can be reproduced.  Again, however, 
it is not immediately obvious how MMRs of widely different slopes might be understood
physically in this picture, or indeed why more massive GCs formed preferentially from
regions of more enriched dense gas.

Does NGC 6166 display this phenomenon?  
A simple approach would be to divide the sample at $(F475W-F814W) = 1.55$
where the two components cross (Fig.~\ref{fig:chisto}) and then derive the trend of
color with luminosity without magnitude-binning   
(though this approach will tend to smooth over any smaller-scale variations with luminosity).
Linear solutions give derived slopes 
$\Delta(475-814)/\Delta(mag) = -0.032 \pm 0.004$ (blue) and
$\Delta(475-814)/\Delta(mag) = 0.002 \pm 0.005$ (red), which translate into
$\gamma(\rm blue) = 0.25 \pm 0.03$ and $\gamma(\rm red) = -0.02 \pm 0.04$.  
Quadratic polynomial fits were also tried, but were not visibly different.  
These correlations are valid
for $F814W < 27$ ($M_I \lesssim -8.6$, or $M \gtrsim 2 \times 10^5 M_{\odot}$),
which is near the turnover (peak) of the GCLF.

\begin{figure}[t]
 \vspace{0.0cm}
 \begin{center}
\includegraphics[width=0.5\textwidth]{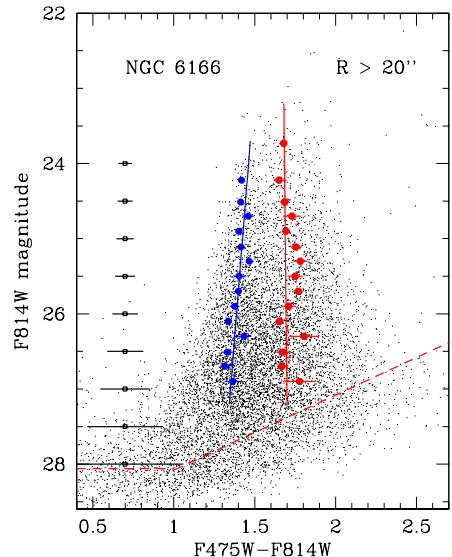}
\end{center}
\vspace{-0.5cm}
\caption{Color-magnitude data for the NGC 6166 GCs, now with the mean GC
color in magnitude bins taken from Table \ref{tab:gmm_mag}.}
\vspace{0.5cm}
\label{fig:cmd_fid}
\end{figure}

A more rigorous approach that would specifically account for the overlap between MP and MR modes, as well
as their different dispersions, is to
define mean points $(\mu_1, \mu_2)$ as a function of magnitude through bimodal-Gaussian fitting
in relatively small magnitude bins. 
The mean points can then be used to define any systematic trend with magnitude.  Mean points in $0.2-$mag intervals
derived this way are superimposed on the CMD in Figure \ref{fig:cmd_fid}.  A linear fit to each set of 
points then gives
$\Delta(475-814)/\Delta(mag) = -0.037 \pm 0.008$ (blue) and
$\Delta(475-814)/\Delta(mag) = 0.005 \pm 0.007$ (red), or 
$\gamma(\rm blue) = 0.29 \pm 0.06$ and $\gamma(\rm red) = -0.04 \pm 0.05$.  The slopes obtained through
either method are closely consistent.  On the MR sequence, in rough terms there is no significant
change in mean color over almost 4 magnitudes in luminosity:  but in more detail, 
the trend in color is not a simple one and may not even be monotonic.  

Along the MP sequence, however, a more consistent signal shows up     
indicating that a modest MMR is present.  
The slope $\gamma \simeq 0.25-0.3$ is near the middle of 
the range seen in other large ellipticals \citep{harris09a,cockcroft_etal2009,peng_etal2009,fensch_etal2014}.
A particularly good comparison is with M87, where the blue sequence has
$\gamma = 0.25 \pm 0.05$ for the luminosity range $M_I \lesssim -9$ \citep{harris09b,peng_etal2009},
very similar to what we find here.
Notably however, in both these galaxies the slope $\gamma$ shows no indication of increasing with luminosity. 
A constant $\gamma$ is inconsistent with the basic self-enrichment model \citep{bailin_harris09}, which requires
curvature in the MP sequence starting with a near-vertical base below about $10^6 M_{\odot}$ 
\citep[see also][for other quantitative examples]{mieske_etal2010}.

We note that for $F814W < 24.0$ (top panel of Fig.~\ref{fig:chisto_mag}), 
the blue sequence fades out but the red sequence continues to still higher luminosity.
In this magnitude range, a unimodal Gaussian function for the color distribution can be rejected at only
the 92\% significance level, whereas in all other luminosity bins a single Gaussian is
rejected at $> 99$\% confidence.
$F814W \simeq I < 24.0$ corresponds to $L$ $\gtrsim 1.9 \times 10^6 L_{\odot}$
or a mass range $M \gtrsim 3.7 \times 10^6 M_{\odot}$ for $(M/L) \simeq 2$ \citep{mclaughlin_vandermarel05}.
As seen in Fig.~\ref{fig:cmd4}, most of these high-luminosity red GCs are in the inner
$\simeq 50$ kpc.  A very similar trend for the red sequence to reach higher is seen in 
some other BCGs such as NGC 3311 \citep{wehner_harris2007} and NGC 4874
\citep{harris_etal09}, though other BCGs do not show it and thus
it does not seem to be universal.  

A potentially connected observation is that dwarf galaxies, which contain
primarily blue MP clusters, also have GCLFs that are narrower and less extended to high luminosities
than in giant galaxies \citep{villegas2010}.  Thus, any part of the GC population accreted at later
times from dwarf satellites would have added to the blue GC total but would not have added ones 
at the highest luminosities.  In the context of formation models \citep{li_gnedin14}
larger galaxies contribute more clusters,        
with higher metallicity. They populate the cluster luminosity function to higher luminosities, and therefore, the    
brightest clusters are expected on average to be red. 

The MMR phenomenon makes it clear that there is much we do not yet understand about the formation processes
and internal enrichment histories of massive star clusters, particularly in the regime above $\sim 10^6 M_{\odot}$.  
Additional physically motivated theory is still needed to fully explain the MMR patterns in 
different galaxies.

\begin{figure}[t]
 \vspace{0.0cm}
 \begin{center}
\includegraphics[width=0.5\textwidth]{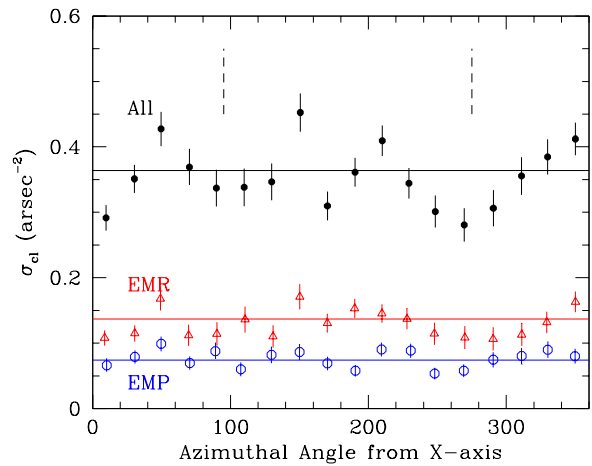}
\end{center}
\vspace{-0.5cm}
\caption{Assessment of the distribution of GCs around NGC 6166 by azimuthal angle $\theta$.
	The density (number per arcsec$^{-2}$) of GCs in the magnitude range $F814W < 27.2$ and in the radial range $20'' < R < 70$
is plotted in $20^o$ sectors.  \emph{Black solid points:} GCs of all metallicities.  
\emph{Red triangles:} Extreme metal-rich (EMR) GCs.  \emph{Blue circles:} Extreme metal-poor
(EMP) clusters.  Mean values for each sample are shown as the horizontal lines.  The vertical
dashed lines at top mark the orientation angle of the \emph{isophotal minor axis}; if the GCS
follows the same orientation as the halo light, then the cluster counts should be lowest at those angles.}
\vspace{0.5cm}
\label{fig:azimuth}
\end{figure}

\section{The Spatial Distribution: Halo Light Versus GCS}

The GCS around this supergiant galaxy is clearly very extended, continuing out well past the outer boundary
of the ACS/WFC field and on through the WFC3 field.  B15 showed that the same is true for the integrated
halo light.  How well do these two types of stellar halo populations match up, and is the correlation
affected by GC metallicity?

Evaluating the effects of metallicity is made difficult by the heavy overlap between MP and MR components.
To help isolate the metallicity trends more clearly, we therefore define two GC subsamples:  an ``Extreme Metal-Poor" (EMP)
sample with color indices bluer than the peak of the MP component ($(F475W-F814W) < 1.401$); and an
``Extreme Metal-Rich'' (EMR) sample with colors redder than the peak of the MR component
($(F475W-F814W) > 1.715$).  This culling guarantees that the EMP component is minimally contaminated
by overlap with the MR clusters, and vice versa for the EMR component \citep[see][for a similar treatment
of NGC 1399 and M87]{forte_etal2007}.
As well as testing the entire GC system, we can then determine the azimuthal and radial distributions
of its extreme low- and high-metallicity components.

\subsection{Azimuthal Dependence}

The surface brightness (SB) profile of the NGC 6166 stellar halo
changes shape significantly with radius (B15); it is nearly
round in the inner halo but elongates to an ellipticity $1 - (b/a) = \epsilon \lesssim 0.5$ at the outermost radii of their data.
However, the position angle of the isophotal major axis stays nearly constant at $\sim 30^o$ E of N.
Before attempting to match up the radial distribution of the halo light with the GC counts, we should
therefore determine if they have similar \emph{azimuthal} distributions.

In the range $R = 20''-70''$ ($12 - 45$ kpc) we can work with a sample of GC counts that is azimuthally complete (i.e. 
completely enclosed in the ACS field; though we do not make the second-order corrections for the small gap
between the two ACS detectors or the $10''$ exclusion circle around the satellite galaxy).  
In Figure \ref{fig:azimuth}, the number density of GCs in this radial range and with $F814W < 27.2$
is shown plotted versus position angle $\theta$ relative to the x-axis of the ACS field.  The counts
are made in $20^o$ sectors.  The translation between $\theta$ and the fiducial directions on the sky is that
North lies $28^o$ clockwise from (i.e. below) the x-axis and East lies $28^o$ clockwise from
the y-axis.  By using the iterative method of moments discussed by \citet{mclaughlin_etal1994},
we find the following results:
\begin{enumerate}
	\item{} For all GCs combined, the mean ellipticity is $\epsilon = 0.26 \pm 0.06$ with major axis at
		$(45 \pm 11)^o$ E of N.
	\item{} For the EMP GCs, $\epsilon = 0.17 \pm 0.13$ with major axis at $(63 \pm 74)^o$ E of N.  
		As is also evident from Fig.~\ref{fig:azimuth}, both parameters are weakly determined and
		the assumption that the EMP cluster distribution is intrinsically spherical cannot be
		clearly rejected.
	\item{} For the EMR GCs, $\epsilon = 0.33 \pm 0.11$ with major axis at $(42 \pm 10)^o$ E of N. 
\end{enumerate}

These results indicate that the azimuthal shape of the GCS depends on metallicity.
Over the same radial range $20''-70''$, the halo light ellipticity increases from $\epsilon \simeq 0.25$ to 0.37,
and is oriented $\simeq 33^o$ E of N with only $\pm 2^o$ variation.  The azimuthal parameters of the halo light
thus closely resemble those of the EMR clusters, but not the EMP clusters.

\subsection{Radial Dependence}

Isophotal contours for galaxy halos are routinely measured by ellipse fitting.  By contrast, GC counts
are usually done in \emph{circular} annuli. Any noncircularity  in the GC distribution
can be clearly gauged only for cases of extremely elliptical distributions, or for galaxies like BCGs
where the statistical sample of GCs is very large. Even so, is difficult to calculate both radial and azimuthal
parameters in fine radial steps
as is done for the halo light \citep{mclaughlin_etal1994}.  

This issue is not of major importance for galaxies
in which both GCS and halo light have small ellipticities.  But in NGC 6166, the halo light
becomes quite elongated at large radius, so 
we correct the SB profile $\mu_V$ 
back to an equivalent circular form that can then be directly matched to the GCS profile.  
For each elliptical annulus $a$ 
for which $\mu_V$ and $e$ are tabulated (as given in Table 3 of B15), 
we then calculate the radius $R_{eq} = \sqrt{ab}$ of the \emph{circular} annulus that has the \emph{same mean surface brightness}
averaged around the circle, i.e. ${\langle \mu_V(R_{eq},\theta) \rangle}_{\theta} = \mu_V(a)$
where $\theta$ is the azimuthal angle of any point on the circle
\citep[see, e.g.,][]{carter1978,bender_etal1988,mclaughlin_etal1994}. 

\begin{figure}[t]
 \vspace{0.0cm}
 \begin{center}
\includegraphics[width=0.5\textwidth]{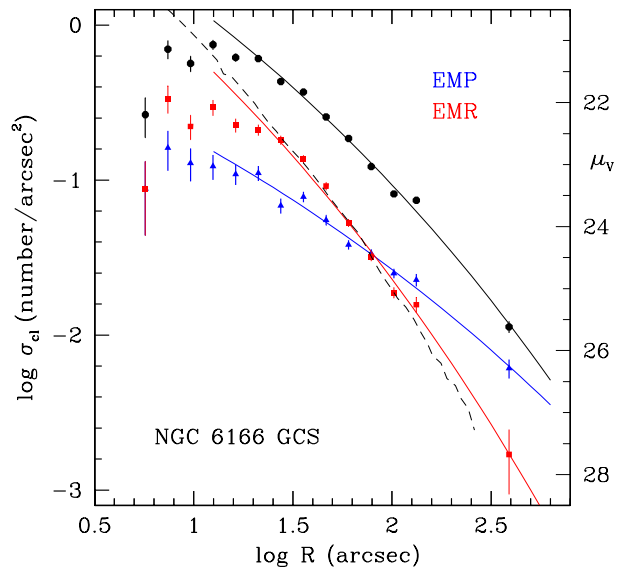}
\end{center}
\vspace{-0.5cm}
\caption{Radial profiles for the globular cluster system around NGC 6166, and for the 
	EMP and EMR subsystems defined in the text.  
	The data are plotted in log-log form in units of number of GCs per arcsec$^2$ versus projected
	radius $R$.  GCs in the magnitude range $23 < F814W < 27$ are included in the totals.
	\emph{Solid dots and line} are the data for all GCs combined.
	\emph{Blue triangles and line} are the extreme metal-poor GCs.   
	\emph{Red squares and line} are the extreme metal-rich GCs.
\emph{Black dashed line:}  Integrated surface brightness profile $\mu_V(R)$ for the NGC 6166 halo,
shifted vertically to align with the EMR GCs.}
\vspace{0.5cm}
\label{fig:radprof}
\end{figure}

Figure \ref{fig:radprof} shows this circularly-adjusted $\mu_V$ profile in comparison with the
number density of all GCs, $\sigma_{cl}(R)$.  The GC sample includes those in the magnitude range $23 < F814W < 27$,
over which the photometric completeness
is high for both ACS and WFC3.
As before, we have assumed zero field
contamination (see Section 5).  The outermost datapoint in Fig.~\ref{fig:radprof} is the value for the
entire WFC3 comparison field.  The five innermost points all lie within
$R \lesssim 20''$, for which the measurements become progressively less complete and less certain.

For $R > 20''$ where the completeness is high, a simple power-law decline does not accurately match
the shape of the GCS profile.  
Instead, we try a S\'ersic-type function in its classic form \citep{sersic1968},
\begin{equation}
	\sigma_{cl} \, = \, \sigma_e {\rm exp}(-b_n [({R \over R_e})^{1/n} - 1] )
\end{equation}
where $\sigma_{cl}$, the number of GCs per unit area, replaces the usual surface brightness
$I(R)$, $R_e$ is the effective radius enclosing half the population, $n$ is an index giving the
steepness of falloff of the profile, and $b_n \simeq 1.992 n - 0.3271$ \citep{caon_etal1993,graham_driver2005}.  
Using the datapoints for $R > 20''$, we solve for the free parameters
($n, R_e, \sigma_e$) by weighted $\chi^2$ minimization.
For the total GC population (solid black circles 
in Fig.~\ref{fig:radprof}) we find a best-fit $n = 6.7$ (solid black line in the Figure),
although the $\chi^2$ minimum is a shallow one and any $n-$values in 
the range $\sim 6-7$ provide good fits.  Plainly, however, the GCS defines a shallower
distribution than the halo light (shown as the dashed line).

There is no clear transition in the GCS profile to the ICM; or, if there
is, it lies further out than $R \sim 300$ kpc.  Similarly, B15 conclude that ``... the cD halo
is not distinguishable using photometry alone''. This feature is in contrast to the Coma
cluster, where \citet{peng_etal11} found the GCS profile in NGC 4874 to become rather suddenly flatter
beyond $R \gtrsim 150$ kpc. At larger radii the Coma IGC population
dominates, adding up to perhaps twice as many GCs as ones belonging to the central galaxy.
Interestingly however, \citet{peng_etal11} also find that in the IGC population, MP GCs outnumber
MR ones by 4:1, not unlike the ratio we find here for our outer WFC3 field (Table \ref{tab:gmm_rad}).

In Figure \ref{fig:radprof}, we also show the EMP and EMR 
subsamples of clusters separately. 
S\'ersic fits to these components give $n(EMP) \simeq 7.6$, $n(EMR) \simeq 6.8$.  
The EMP population is strikingly more extended than the EMR one.  For purposes of rough comparison, in power-law form
where $\sigma_{cl} \sim R^{\alpha}$, we find $\alpha \simeq -1.0$ (EMP) but $\alpha \simeq -1.8$ (EMR).
The shallow MP slope is close to an isothermal profile that would characterize a dark-matter halo.
The crossover radius where $\sigma(EMP) = \sigma(EMR)$ is at $R \simeq 50$ kpc.

Notably, the radial profile for the EMR population tracks the profile for the \emph{halo light} 
much more closely.  The adjusted halo light profile is shown as
the dashed line, normalized to the EMR GCS.  
The normalization factor is that 1 
\emph{metal-rich} GC brighter than $F814W = 27$ is equivalent to a halo luminosity $V = 21.95$ (or $M_V = -13.6$).
For the entire range $R > 20''$, little or no significant
difference can be seen between them.  The outermost datapoint for the EMR sample can be seen to lie somewhat
above the outward extrapolation of $\mu_V(R_{eq})$, but it is not clear how much weight should be put on it.
That datapoint comes from only the WFC3 field and thus belongs to a very small range $\Delta \theta$ of azimuthal
angle, so it is difficult to define a valid $R_{eq}$ for that radius, particularly because we also do not 
know the axial ratio of the GCS there.

To this result we can add the observation by B15 that the integrated $V-I$ color of the halo
gradually decreases with radius, by 0.1 mag out to $R = 80''$ (50 kpc). 
Notably, a plot of $\mu_I$ if extrapolated outward would look slightly
shallower than $\mu_V$ and would bring the agreement with $\sigma_{cl}(EMR)$ even closer.

In their discussion of the NGC 6166 halo light, B15 find that by a radius $R \simeq 70'' \simeq 45$ kpc, the
halo velocity dispersion has risen to a value $\simeq 800$ km s$^{-1}$ comparable with the A2199
cluster galaxies, suggesting that the cD halo component has become dominant by that point relative to the
core of the galaxy.  They also find that the halo metallicity is $\alpha-$enhanced
([$\alpha$/Fe] = 0.3) out to $R = 59''$, indicating fast star formation within $\sim 1$ Gyr and rapid quenching
after that.  The GCS data provide a way to extend this argument further out.  \emph{If} the 
ratio of MP to MR \emph{halo stars} were to change with radius in the same way as the MP and MR GCs do, then the light
profile in Fig.~\ref{fig:radprof} would follow the total GC population, not the MR component.  
This comparison suggests that the stellar halo of NGC 6166 -- both cD and core components -- remains
moderately metal-rich even at large radius.

The question these comparisons leave us with is the origin of the many thousands of 
\emph{metal-poor} clusters at large radius.  The major
possibilities are that (a) the MP GCs originated at a very early stage of evolution in 
the many small, metal-poor halos just beginning their
star formation, at a redshift when they still followed the shallow dark matter halo profile; or
(b) many of them are from a later 
accreted population of disrupted small satellite galaxies or the outer halos of larger galaxies, 
the majority of which would be metal-poor GCs.
Both factors can be part of the story.  In either case, the argument relies on the empirical result
that the GC specific frequency (number of GCs per unit halo light) \emph{increases} dramatically as
metallicity \emph{decreases}. That is, low-metallicity environments were much more efficient at forming
massive star clusters in the early universe 
\citep{harris_harris02,forte_etal2005,harris_etal2007,kruijssen2014,forte_etal2014,peacock_etal2015}.  
Accreted stellar populations dominated by dwarfs
could then add GCs with only minor effects on the metal-rich halo light component.

The synthesis to be drawn from the combined data is largely in agreement with the conclusions of
B15 from their surface photometry and integrated-light spectroscopy, that 
the extended cD-type halo of NGC 6166
consists of tidal debris from other galaxies in the cluster.  These other galaxies were likely 
themselves to have a wide variety of halo metallicity gradients and dispersion \citep{monachesi_etal2015}.
In the dense central environment of A2199
star formation proceeded intensely and rapidly for a short period of time, with later additions to
the cD halo coming from dynamical disruption processes.

Finally, we estimate the total GC population and specific frequency $S_N$ \citep{harris_vandenbergh1981}.
The local specific frequency will increase outward since $\sigma(GCS)$ is shallower than the halo light
profile, so we restrict the calculation to the outermost radius to which either component has been traced,
namely $R \simeq 415''$ = 260 kpc.  Integrating the GCS S\'ersic profile from $20'' - 415''$ gives
$N_{GC} = 18600 \pm 1000$ brighter than $F814W = 27.0$. To this we add $\simeq 900$ more for 
$R < 20''$, taking the $\sigma_{cl}$ value at $20''$ and assuming conservatively that it is constant further in.
Since the limiting magnitude is almost exactly at the turnover (peak frequency) of the GCLF, this total
is then doubled to account for fainter GCs, giving $N_{GC}(tot) = 39000 \pm 2000$.
To the same outer radius, B15 calculate a total integrated $V$ magnitude $V(tot) = 11.75$ or
$M_V^T = -23.85$.  The global specific frequency is then
\begin{equation}
	S_N \, = \, N_{GC}(tot) \cdot 10^{0.4(M_V^T + 15)} \, = \, 11.2 \pm 0.6.
\end{equation}
A $S_N$ this large is quite similar to the values found for NGC 4874, M87, and other BCGs
\citep{harris_etal1995,peng_etal11,harris09b}.
For a mean GC mass of $\sim 2 \times 10^5 M_{\odot}$, the total mass fraction of
GCs to galaxy stellar mass in NGC 6166 is roughly $M(GCS)/M_{\star} \simeq 5 \times 10^{-3}$.

Integrated outward to the same limiting radius $R(max) = 415''$, the total numbers of
blue and red GCs are $N(MP) = 22300 \pm 1500$ and $N(MR) = 16700 \pm 1400$.
The global ratio $N(blue)/N(tot)$ is then $0.57 \pm 0.05$.

\begin{figure}[t]
 \vspace{0.0cm}
 \begin{center}
\includegraphics[width=0.5\textwidth]{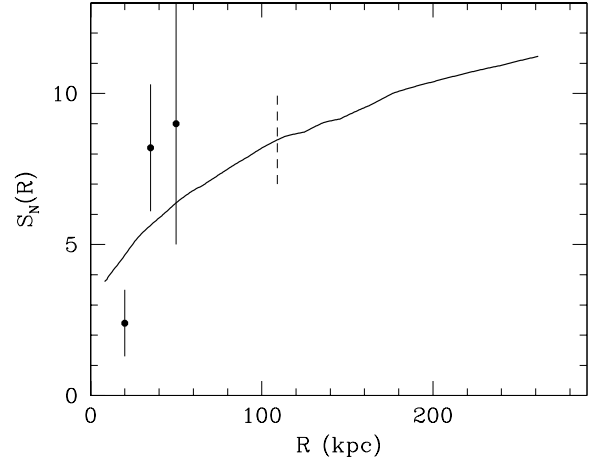}
\end{center}
\vspace{-0.5cm}
\caption{Increase of specific frequency $S_N$ with radius around NGC 6166.  Here, $S_N(R)$
is defined as the total cluster population within radius $R$, divided by the total $V-$band
luminosity within $R$. The vertical dashed line shows the outer boundary for our ACS/WFC field;
at larger radii the curve relies on the WFC3 data and the fitted S\'ersic profile
for the GCS (see text).  The three points with errorbars indicate estimates in the earlier
literature from ground-based imaging.}
\vspace{0.5cm}
\label{fig:sntot}
\end{figure}

Because the overall GCS surface density profile $\sigma_{cl}(R)$ is shallower than the 
surface brightness profile $I_V(R)$ (Fig.~\ref{fig:radprof}), $S_N$ increases outward.  The
quantitative trend is shown in Figure \ref{fig:sntot}, obtained by integrating the appropriate
S\'ersic profiles for the GCS and for the $V-$band surface brightness profile from B15. 
For the inner halo $R \lesssim$ 40 kpc,
the specific frequency is at a level $S_N \sim 5$ that is in the normal range for large early-type
galaxies, but it rises smoothly outward to the limit of our data.  

Comparisons can be made with the older
published estimates obtained from ground-based imaging.
\citet{pritchet_harris1990} found $S_N \simeq 2.4 \pm 1.1$ within $R \sim 20$ kpc, while we
obtain $S_N = 4.1$ to that radius.  Because they used the outer parts of their $\sim 100''$ field
of view to define background, it is now clear that the GCS population was oversubtracted.
\citet{bridges_etal1996} were able to define the background count level with a much more
remote control field and found $S_N = 9 (+9,-4)$ to within $\simeq 50$ kpc, whereas our value
is $S_N(50 kpc) = 6.5$, within their estimated range.  Lastly,
\citet{blakeslee_etal1997} determined $S_N = 8.2 \pm 2.1$ within $R \simeq 35$ kpc through
a combination of resolved-object photometry and surface brightness fluctuation measurement; 
by comparison we obtain $S_N(35 kpc) = 5.6$.  Thus these earlier estimates 
have accuracies of typically a factor of two.

\section{Summary}

We have presented the first comprehensive photometric study of the extraordinarily rich
globular cluster system around NGC 6166, the BCG in A2199 and a classic cD-type galaxy.
Two-color photometry from the HST ACS and WFC3 cameras was used to measure the GC population
with a limiting magnitude \emph{at} the GCLF turnover point.
Our principal findings are these:
\begin{enumerate}
\item{} The GCS is extremely populous, easily detectable
	out to at least 260 kpc and totalling 39,000 GCs to that radius.  
	The global specific frequency to that radius is $S_N = 11.2 \pm 0.6$.
\item{} The metallicity distribution of the GCs can still be very well described by a bimodal
	Gaussian-type function, and these two modes are separated
	by $\Delta$[Fe/H] = 1.0 dex as is the norm for other galaxies.
	But the metal-rich and metal-poor modes both have larger dispersions in NGC 6166 than
	in other systems, making the two modes overlap significantly and filling the 
	usually-sparse intermediate-metallicity zone.  Both these features are reminiscent of
	the recent \citet{li_gnedin14} models in which GC formation is driven at every stage
	by halo mergers.  With NGC 6166, we may be seeing
	the results of a hierarchical formation process so extended and complex that the simplistic
	bimodal, two-phase scenario is no longer an effective picture of its history.
	This regime of extreme galaxy mass and environmental richness is essentially populated only by BCGs.
\item{} The GCS shows a strong global metallicity gradient, but this results almost entirely 
	from the decreasing ratio of MR to MP clusters with increasing radius.  The mean metallicity
	of each mode changes little with radius.
\item{} The radial profile of the $V-$band \emph{halo light} matches the \emph{metal-rich} GCs 
	extremely well for all radii $R > 20''$ (12 kpc), falling roughly as $R^{-1.8}$ in surface
	intensity ($\mu_V$) or GC number density ($\sigma_{gc}$).  By contrast, the metal-poor GCs
	follow a much shallower profile as $\sigma \sim R^{-1.0}$, more nearly matching an isothermal
	dark-matter halo.  The red, metal-rich GCs lie in an elliptical spatial distribution that also
	matches the shape of the halo light.
\item{} The blue GC sequence shows a modest mass-metallicity relation where heavy-element abundance
	increases with cluster mass as $Z \sim M^{0.25}$.  But at the highest GC luminosities
	($\gtrsim 2 \times 10^6 L_{\odot}$) the red sequence reaches higher and the MDF becomes unimodal.
	No single physical model is yet able to account satisfactorily for the puzzling variety of
	MMR structures that have already been seen in large galaxies.
\item{} We find no clear spatial transition between the inner core galaxy and its cD envelope, or the
	ICM.  In this respect it behaves the same way as does the halo light profile, but differs
	from the more abrupt transition seen in the Coma cluster and its BCG, NGC 4874.
	\end{enumerate}

\section*{Acknowledgements}

Based on observations made with the NASA/ESA Hubble Space Telescope, obtained at the Space Telescope Science Institute, 
which is operated by the Association of Universities for Research in Astronomy, Inc., under NASA contract NAS 5-26555. 
WEH acknowledges financial support from NSERC (Natural Sciences and Engineering Research Council of Canada).
BCW acknowledges support from NASA grant HST-GO-12238.001-A. 
OG was supported in part by NASA through grant NNX12AG44G, and by NSF through grant 1412144.
DG gratefully acknowledges support from the Chilean BASAL Centro de
Excelencia en Astrof\'isica y Tecnolog\'ias Afines (CATA) grant PFB-06/2007.  

{\it Facilities:} \facility{HST (ACS, WFC3)}

\appendix

\section{MDF Parameters for the Milky Way}

\begin{figure}[t]
 \vspace{0.0cm}
  \begin{center}
	  \includegraphics[width=0.5\textwidth]{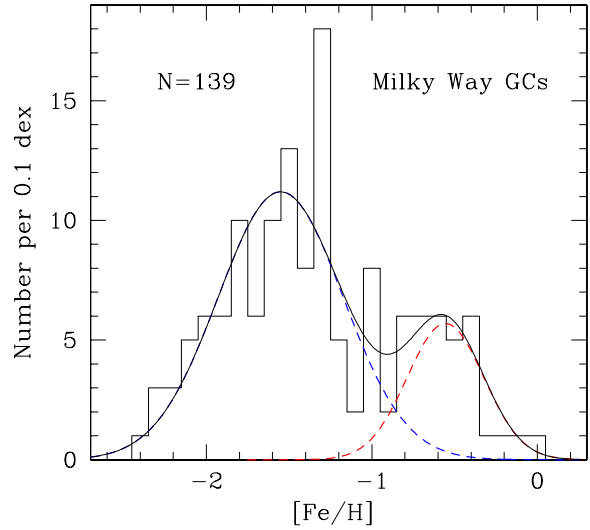}
  \end{center}
  \vspace{-0.5cm}
  \caption{Metallicity distribution function for the GCs in the Milky Way.
	  The MP (blue dashed line) and MR (red dashed line) component solutions
  are shown in the figure and listed in the text.}
  \vspace{0.5cm}
  \label{fig:milkyway}
  \end{figure}

Although the Milky Way is far from being a BCG, it represents the foundation of the
bimodality paradigm for globular cluster systems.  To provide a comparison for the BCGs
listed in Table \ref{tab:mdf_bcg},
we show the Milky Way cluster metallicities
in Figure \ref{fig:milkyway}, based on the most recent compendium of measurements.  Here, 139
clusters with reddenings $E_{B-V} < 1.6$ are shown, with [Fe/H] values from \citet{harris96} (2010 edition).
A unique advantage of this sample is that
the great majority of these metallicities are determined directly from high dispersion spectroscopy 
of the cluster stars.

The fitted GMM-derived parameters, this time directly in units of [Fe/H] rather than color, are 
$\mu_1 = -1.55 \pm 0.07$, $\mu_2 = -0.55 \pm 0.10$, dispersions $\sigma_1 = 0.38 \pm 0.04$ dex,
$\sigma_2 = 0.23 \pm 0.05$ dex, and MP fraction $p_1 = 0.69 \pm 0.09$.
The effect of small-sample
statistical scatter is evident in the bin-to-bin differences, but the biggest single difference between this and
the BCGs is perhaps the much smaller dispersion $\sigma_2$ for the MR clusters.  Once again, the results are
suggestive of a more prolonged and dominant metal-rich formation mode in the biggest galaxies.

Interestingly, the width of the MP component (0.38 dex) is larger than for most of the BCGs listed
above.  This spread
causes several clusters that nominally belong to the MP component
to fall in the intermediate-metallicity zone [Fe/H] $\sim -1$.
Given the small sample size, it is risky to place too much significance on this feature, but for
completeness we therefore also ran a trimodal solution with GMM.  This model fit gives
component means $\mu_1 = -1.66 \pm 0.22$, $\mu_2 = -1.30 \pm 0.39$, $\mu_3 = -0.63 \pm 0.11$;
dispersions $\sigma_1 = 0.34 \pm 0.10$, $\sigma_2 = 0.08 \pm 0.10$, $\sigma_3 = 0.27 \pm0.07$;
and fractions $p_1 = 0.60 \pm 0.25$, $p_2 = 0.11 \pm 0.23$, $p_3 = 0.29 \pm 0.08$.
The extra intermediate-metallicity mode is quite weakly determined, and from goodness-of-fit ($\chi^2/\nu$) criteria
a bimodal fit is preferred over a trimodal one.  

For the Milky Way clusters much other information is available concerning the cluster kinematics and
spatial distributions in three dimensions, and even grouping of subsets into streams associated with
tidally disrupted satellites.  In most other galaxies this level of detail is unavailable, but it is
encouraging that the more basic features of the GCS that can be measured in other galaxies -- the 
MDF and projected spatial distribution -- correlate well with the more detailed characteristics of
the GCS components in the Milky Way.

\makeatletter\@chicagotrue\makeatother

\bibliographystyle{apj}
%\bibliography{gc}

\label{lastpage}

\end{document}